\documentclass[fleqn]{2020SCGE}
\usepackage{color}
\usepackage{float}
\usepackage{natbib}
\usepackage{gensymb}
\usepackage{graphicx}
\usepackage{txfonts}
\usepackage{longtable}
\usepackage{url}
\usepackage{hyperref}%PDF
\begin{document}
%\linenumbers

\ensubject{subject}

%%%%%%%%%%%%%%%%%%%%%%%%%%%%%%%%%%%%%%%%%%%%%%%%%%%%%%%
%%% Authors do not modify the information below

%Letter to the Editor
\ArticleType{Article}
\SpecialTopic{Special Topic: Peering into the Milky Way by FAST}
\Year{2022}
\Month{December}
\Vol{65}
\No{12}
\DOI{10.1007/s11433-022-2033-2}
\ArtNo{129704}
\ReceiveDate{July 14, 2022}
\AcceptDate{November 10, 2022}
\OnlineDate{November 23, 2022}
%%%%%%%%%%%%%%%%%%%%%%%%%%%%%%%%%%%%%%%%%%%%%%%%%%%%%%%

\title{ Peering into the Milky Way by FAST: 
\\ III. Magnetic fields in the Galactic halo and farther spiral arms revealed by the Faraday effect of faint pulsars
\\ }{Peering into the Milky Way by FAST: III. Magnetic fields in the Galactic halo and farther spiral arms revealed by the Faraday effect of faint pulsars}

%%% Corresponding author: 
%%%   \author[number]{Full name}{{email@xxx.com}}
%%% General author: 
%%%   \author[number]{Full name}{}
%\author[]{{\color{red}{(The author names, affiliations and ranking will be finalized after the Fermi collaboration review.)}}}{}

\author[1*]{Jun Xu }{}%$^{\dag}$}{}
\author[1,2*]{JinLin Han}{} %$^{\ddag}$}{}
\author[1,2]{PengFei Wang}{}
\author[1,2]{Yi Yan}{}
%\author[1]{.........?}{}
%
\footnote{Corresponding authors (Jun Xu email: xujun@nao.cas.cn; JinLin Han email: hjl@nao.cas.cn)}

%%% Author information for page head.
\AuthorMark{J. Xu} % et al.: Galactic magnetic fields.}

%%% Authors for citation. 
  \AuthorCitation{J. Xu, J. L. Han, P. F. Wang, Y. Yan}

%%% Address.
%%% \address[number]{Address, City {\rm Postcode}, Country}
\address[{\rm1}]{National Astronomical Observatories, Chinese Academy of
     Sciences, Beijing 100101, China}
\address[{\rm2}]{School of Astronomy, University of Chinese Academy of Sciences, Beijing 100049, China}     

%%% Abstract.
\abstract{
The Five-hundred-meter Aperture Spherical radio Telescope (FAST) is the most sensitive radio telescope for pulsar observations. We make polarimetric measurements of a large number of faint and distant pulsars using the FAST. We present the new measurements of Faraday rotation for 134 faint pulsars in the Galactic halo. Significant improvements are also made for some basic pulsar parameters for 15 of them. We analyse the newly determined rotation measures (RMs) for the Galactic magnetic fields by using these 134 halo pulsars, together with previously available RMs for pulsars and extragalactic radio sources and also the newly determined RMs for another 311 faint pulsars which are either newly discovered in the project of the Galactic Plane Pulsar Snapshot (GPPS) survey or previously known pulsars without RMs. The RM tomographic analysis in the first Galactic quadrant gives roughly the same field strength of around 2~$\mu$G for the large-scale toroidal halo magnetic fields. The scale height of the halo magnetic fields is found to be at least 2.7$\pm$0.3~kpc. The RM differentiation of a large number of pulsars in the Galactic disk in the Galactic longitude range of $26^{\circ}<l<90^{\circ}$ gives evidence for the clockwise magnetic fields (viewed from the north Galactic pole) in two interarm regions inside the Scutum arm and between the Scutum and Sagittarius arm, and the clockwise fields in the Local-Perseus interarm region and field reversals in the Perseus arm and beyond.
}%
%%% Keywords.


\maketitle

%%%\tableofcontents%

\begin{multicols}{2}

%%%%%%%%%%%%%%%%%%%%%%%%%%%%%%%%%%%%%%%%%%%%%%%%%%%%%%%%%%%%
%% Text of article.
%%%%%%%%%%%%%%%%%%%%%%%%%%%%%%%%%%%%%%%%%%%%%%%%%%%%%%%%%%%%
\section{Introduction}           
\label{intro}

The Five-hundred-meter Aperture Spherical radio Telescope \citep[FAST,][]{nan08,nlj+11}, mounted with the 19-beam L-band receiver, is an extremely sensitive radio telescope to observe pulsars and spectral lines for HI gas and ionised gas. The Galactic Plane Pulsar Snapshot (GPPS) survey\footnote{ http://zmtt.bao.ac.cn/GPPS/} is to search for pulsars in the FAST accessible sky within $|b| \leq 10^\circ$ \citep{hww+21}. During the GPPS survey observations, the piggyback spectral data are simultaneously recorded  with 1024~K channels for the band of $1000 - 1500$ MHz. 
This series of papers are dedicated to the investigations of the interstellar medium by FAST. The first paper of this series by \citet{hhh+22a} is on the exquisite HI structures from the high resolution and high sensitivity piggyback HI line observations of the FAST GPPS survey. 
The second paper of this series by \citet{hhh+22b} 
is on the piggyback recombination line observations of the FAST GPPS survey for HII regions and interstellar ionised gas. This is the third paper on the interstellar magnetic fields revealed by measuring Faraday effect of large number of weak pulsars, for which the polarization properties are difficult to measure without FAST. The fourth paper of this series by \citet{grs+22} is using the FAST 19-beam L-band receiver of the FAST to scan a sky area of  $5^{\circ}\times7^{\circ}$ to get the image of radio continuum emission and identify two large nearby supernova remnants G203.1+6.6 and G206.7+6.0, one of which is very close to our Sun. 

Magnetic fields are ubiquitous in galaxies and play crucial roles in astrophysics and astroparticle physics. For example, the magnetic fields affect the evolution of molecular clouds and star formation \citep[see reviews e.g.][]{cru12} and are agents for transport of cosmic rays \citep[e.g.][]{ps03,ls11} and propagation of axionlike particles \citep[e.g.][]{xpg+21}. However, many aspects of the Galactic magnetic fields are unknown because the fields have merely been measured for some components in limited regions \citep[see a review by][]{han17}. It is very challenging to reveal the global magnetic structure of the Milky Way Galaxy since we live inside at nearly the edge of the disk.

Several tracers for interstellar magnetic fields have been largely observed, including starlight polarization \citep[e.g.][]{hei96,cppt12}, polarized thermal emission from aligned dust grains in molecular clouds \citep[e.g.][]{paa+16b}, synchrotron emission from the diffuse interstellar medium (ISM) \citep[e.g.][]{blw+13,paa+16}, Zeeman splitting of spectral lines from clouds or clumps \citep[e.g.][]{cwh+10}, and Faraday rotation of polarized radio sources
\citep[e.g.][]{sk80,tss09}. Generally, these tracers only give information for one component of the magnetic fields either parallel or perpendicular to the line of sight. 

Faraday rotation of linearly polarized emission from pulsars and extragalactic radio sources (EGRS) is an excellent probe for magnetic fields in our Galaxy \citep{han17}. Pulsars are often highly polarized and have negligible intrinsic Faraday rotation \citep{whl11}. They are widely spread throughout the Galaxy at approximately known distances, allowing three-dimensional tomographic mapping of the field structure. The rotation measure (RM) of a pulsar is an integrated quantity of the product of the electron density and magnetic field strength from a pulsar to us, i.e.
\begin{equation}
{\rm RM}=0.812\int_{0}^{D}n_e \mathbf{B} \cdot {\rm d}\mathbf{l},
\end{equation}
where $n_e$ is the electron density in cm$^{-3}$, $\mathbf{B}$ is the vector magnetic field in $\mu$G, $D$ is the distance, and ${\rm d}\mathbf{l}$ is the unit vector along the line of sight towards us in parsecs. The integrated electron density along the line of sight can be measured by the dispersion of pulses, by ${\rm DM}=\int_{0}^{D}n_e {\rm d}l$. The magnetic field component parallel to the line of sight can be directly obtained by 
\begin{equation}
  \left<B_{||}\right>
  =1.232\frac{\rm RM}{\rm DM}
  =\frac{\int_{0}^{D}n_e \mathbf{B} \cdot {\rm d}\mathbf{l}}{\int_{0}^{D}n_e~{\rm d}l},
  \label{meanB1}
\end{equation}
where $B_{||}$ is in $\mu$G, RM and DM are in their usual units (rad~m$^{-2}$ and cm$^{-3}$~pc). When RMs of a large number of pulsars have been measured, the mean field strength between a pair of pulsars at distances $D_1$ and $D_2$ in a given direction can be obtained by
\begin{equation}
  \left<B_{||}\right>_{D_2-D_1}
  =1.232\frac{\rm RM_2-RM_1}{\rm DM_2-DM_1}.
  \label{meanB}
\end{equation}
The reliability of such estimates due to the possible coupling of electron density with a magnetic field in the interstellar medium was questioned by \citet{bssw03}, but has been clarified by detailed simulations by \citet{wkr+09,wkr15} for the diffuse ISM with different Mach numbers. On Galactic scales,  $\left < B_{||} \right >=$ 1.232RM/DM provides a good estimate of the magnetic field along the line of sight \citep{sf21}.

The large-scale Galactic magnetic fields consist of the halo and disc components. The halo fields exist obviously as demonstrated by bright synchrotron sky at low frequencies \citep[e.g.][]{bkb85}. Since we live close to the edge of the Galactic disk, the Galactic halo should be most observable in the mid-latitude regions towards the inner Galaxy. Faraday rotation of a large number of EGRS behind the Galactic halo provides a powerful tool to measure the magnetic fields in the halo. The antisymmetric RM distribution of EGRS in the inner Galaxy found by \citet{hmbb97,hmq99} suggests that the halo fields are probably bi-toroidal with opposite directions above and below the Galactic plane. Such antisymmetry of the RM sky has been confirmed by more and more RM data from large radio continuum surveys \citep[e.g.][]{tss09,scw+19} and collections of published RMs \citep[e.g.][]{xh14,ojr+12,ojg+15,hab+22}. Based on limited knowledge of reversed bi-toroidal fields in the Galactic halo, the global field models have been proposed and developed over the years \citep[e.g.][]{ps03,srwe08,ptkn11,jf12,tf17,xh19}. Analysis of pulsar RMs suggests the scale height of the Galactic halo magnetic field is at least 2.0~kpc \citep{sbg+19} and local halo field strength should be more than 1.6~$\mu$G \citep{xh19}. Analysis of EGRS RMs towards the outer Galaxy indicates different field strengths in the halo above and below the Galactic plane \citep{mmg+12}. There also may exist local vertical magnetic fields \citep[e.g.][]{hq94,mgh+10}.

In the Galactic disk, RMs of pulsars and EGRS have revealed several field reversals between the spiral arms. Early studies of 38 pulsar RMs observed by \citet{man74} show that the local regular magnetic field with a strength of around 2 $\mu$G is directed toward $l\sim 90^{\circ}$, i.e., clockwise when viewed from the north Galactic pole. Later more RMs of pulsars and EGRs support a reversal in the inner Galaxy at or near the Carina-Sagittarius arm \citep[e.g.][]{tn80,sk80,ls89,rk89,cw90}. Moreover, a second reversal
(clockwise) in the inner Galaxy beyond the Crux-Scutum arm is proposed \citep{rl94}, and a further reversal near the Norma arm is identified by \citet{hmq99,hmlq02}. Evidence for field reversal(s) near or exterior to the Perseus arm towards the anticenter is also presented by \citet{ls89,ccsk92,hmq99}. \citet{hml+06} analyzed a larger sample of pulsar RMs in the tangential regions of spiral arms to give clear evidence for large-scale counterclockwise fields in the spiral arms and reversed fields in the interarm regions.
These reversals are supported by independent analysis by \citet{njkk08} and \citet{nk10}. \citet{hmvd18} demonstrated large-scale reversals of the field directions between the arms and the interarm regions more clearly by combining newly observed RMs of more pulsars and more EGRS. 

Some models have been proposed to fit the RM data, such as the ring model \citep{rl94,val05}, the axisymmetric spiral model \citep{val91} and bisymmetric spiral model \citep{hq94,hmq99,id99}, but it is hard to fit just one model for all RM data \citep{mfh08}. The large-scale magnetic fields in the Galactic disk may have a more complex pattern as envisioned  \citep[e.g.][]{obkl17,sls+19,mmob20}. RMs of EGRS behind the Galactic disk show a strong swing between positive and negative, which can be fitted by the field models with only one or two reversals inside the Solar Circle \citep[e.g.][]{bhg+07,vbs+11,jf12,vbo+21}, but cautions should be taken that the integrated RM values over the whole path in the disk from the Galactic outskirts to the Sun are much less sensitive to magnetic field reversals between the arms and the interarm regions along the path. 

Currently, 3341 pulsars have been discovered \citep[see][]{mhth05} \footnote{http://www.atnf.csiro.au/research/pulsar/psrcat/ (version 1.68)}, and 1453 of them have published RMs. 
Among 1045 pulsars in the Galactic halo ($|b|>8^{\circ}$), 472 of them have measured RMs.
The remaining unmeasured pulsars are so faint to be measured efficiently by other radio telescopes, except for FAST. Here we present the new measurements of RMs for 134 pulsars using the L-band 19-beam receiver of FAST in two observation projects from 2020 to 2021. Moreover, \citet{whj+22} have observed a very large sample of pulsars for polarization profiles\footnote{see http://zmtt.bao.ac.cn/psr-fast/}, and determined an additional 311 new RMs for newly-discovered pulsars in the GPPS survey \citep{hww+21} and distant weak known pulsars in the Galactic longitude range $26^{\circ}<l<90^{\circ}$ towards the inner Galaxy. We use these new measurements, together with previously published pulsar RMs and EGRS RMs, to make tomographic analysis of large-scale magnetic field structure in the Galactic halo and disc. The YMW16 electron density model \citep{ymw17} is used to estimate pulsar distances from DMs, which have an uncertainty less than 20 percent for two thirds of pulsars.

The arrangement of our paper is as follows. Our FAST observations and data reduction are described in Section~\ref{obsdata}. We analyse the magnetic field strength and scale height in the Galactic halo in Section~\ref{halofield}, and analyse the magnetic fields in the Galactic disk in Section~\ref{diskfield}. Finally, the conclusions are given in Section~\ref{conclusion}.

\begin{table*}[!t]
\centering
%
%\begin{threeparttable}
\caption{Fundamental parameters of 15 pulsars updated by FAST observations}
\label{parupdate}
\footnotesize
\setlength{\tabcolsep}{4pt}
\renewcommand{\arraystretch}{0.8}
%\scriptsize
\begin{tabular}{lllllllllll} 
\hline
PSR          &  Ref.         & Period  & DM      &  RA(J2000) & Dec(J2000)  & Items     & Period($\sigma$)        & DM($\sigma$)    &   RA(J2000)     &  Dec(J2000)    \\    
name       &              &(s)    &(cm$^{-3}$ pc)&(hh:mm:ss)&(dd:mm:ss)&  updated    &(s)            & (cm$^{-3}$ pc)&(hh:mm:ss)&(dd:mm:ss)    \\
(1)       & (2)          & (3)     &(4)       &(5)       & (6)         & (7)       & (8) &(9)       &(10)         &(11)           \\  
\hline                                                            
J0011+08  &    [1]    & 2.55287 &   24.9  &     00:11:34 & +08:10 & position       &   2.552860(18)     &    25.6(9)  &     00:11:40  &      +08:05  \\
J0146+31  &    [2]    & 0.9381  &   25    &     01:46:15 & +31:04 & position       &   0.938067(4)      &    24.4(5)  &     01:46:41  &      +30:55  \\
J0241+16  &    [3]    & 1.5454  &   16    &     02:41:46 & +16:04 & position       &   1.544897(10)      &    19.3(8)  &     02:41:22  &      +16:04  \\
J0244+14  &    [3]    & 2.1281  &   31    &     02:44:51 & +14:27 & position       &   2.127917(12)    &    29.3(7)  &     02:45:19  &      +14:32  \\
J0711+0931&   [4],[5] & 1.21409 &   46.238&  07:11:36.18 & +09:31:25 & period       &   2.428176(8)     &    46.0(8) &   07:11:36.18 & +09:31:25    \\
J0806+08  &    [3]    & 2.0631  &   46    &     08:06:05 & +08:17 & position       &   2.063092(20)    &    48.4(13)  &     08:06:17  &      +08:12  \\
J0811+37  &   [6],[7] & 1.2483  &   16.95 &     08:11:12 & +37:28 & position       &   1.248281(7)      &    15.9(4)  &     08:11:19  &      +37:31  \\
J0827+53  &    [7]    & 0.0135  &   23.103&     08:27:48 & +53:00 & position       &   0.013526(0)      &    23.11(2)  &     08:28:26  &      +53:04  \\
J0848+16  &    [3]    & 0.4524  &   38    &     08:48:53 & +16:43 & position,period     &   0.904709(2)    &    38.6(3)  &     08:48:43  &      +16:41  \\
J1750+07   &    [1]    & 1.90881 &   55.4  &     17:50:40 & +07:33 & period       &   5.726256(41)     &    55.6(20)  &     17:50:40  &      +07:33  \\
J1802+0128&    [8]    & 0.55426 &   97.97 &  18:02:27.45 & +01:28:23.7 & period     &   1.108528(2)     &    97.8(4) &  18:02:27.45  & +01:28:23.7  \\
J1807+04  &    [3]    & 0.7989  &   53    &     18:07:25 & +04:05 & position       &   0.798766(2)      &    53.0(3)  &     18:07:17  &      +03:59  \\
J1809+17  &    [7]    & 2.0667  &   47.32 &     18:09:06 & +17:04 & position       &   2.066642(12)      &    46.4(7)  &     18:09:07  &      +17:06  \\
J1832+27  &    [6]    & 0.6318  &   46    &     18:32:10 & +27:58 & position       &   0.631707(1)      &    47.5(2)  &     18:32:19  &      +27:49  \\
J1937-00  &    [3]    & 0.2401  &   68.6  &     19:37:09 & --00:17 & position       &   0.2401898(2)      &    67.98(8)  &     19:37:03  &      --00:24  \\

\hline
\end{tabular}
\begin{tablenotes}
\item 
Note: Columns(1)--(6) are names, references, and period, DM, and position in RA(J2000) and Dec(J2000) from the references in Columns(2): [1]: \citet{dsm+16}; [2]: \citet{ttol16};
  [3]: \citet{dsm+13}; [4]: \citet{lzb+00}; [5]: \citet{bkk+16};
  [6]: \citet{ttk+17}; [7]: \citet{scb+19}; [8]: \citet{ebvb01},
  and columns (8)-(11) are parameters obtained in our FAST observations.
  The uncertainty of updated coordinates for columns (10) and (11) is $1'$ in total.
\end{tablenotes}
%\end{threeparttable}
\end{table*}

\begin{figure*}[!t]
\centering \includegraphics[angle=-90,width=0.95\textwidth]{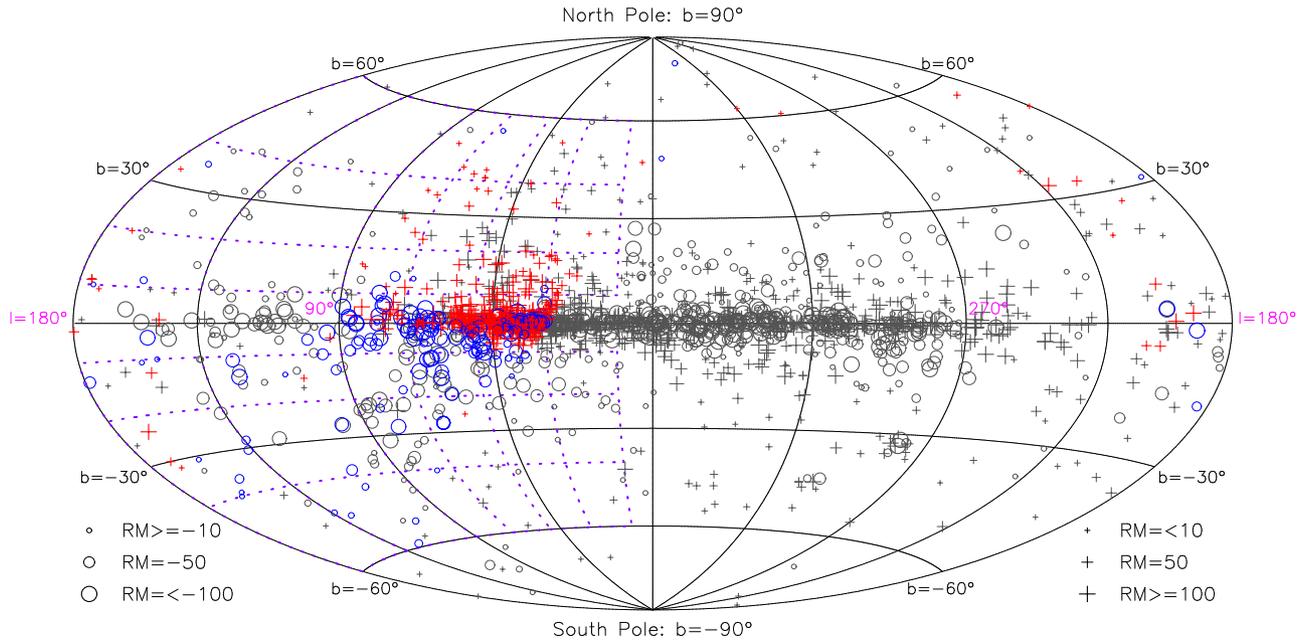}
\caption{The Galactic distribution of RMs for 1860 
  pulsars. The sizes of symbols are proportional to the square root of the RM magnitudes, with limits of 10 and 100 rad~m$^{-2}$. The circles and plus signs represent negative and positive RMs, respectively. The red and blue symbols indicate the new measurements of positive and negative RMs from this work and also from the new FAST pulsar database \citep{whj+22}. Grids divide the sky regions for analysing the magnetic field strength in the Galactic halo in sect.~\ref{halostrength31}. }
\label{psrsky}
\end{figure*}

\begin{table*}[!t]
\centering
\caption{Rotation measures for 134 halo pulsars}
\label{RMtable}
\setlength{\tabcolsep}{4pt}
\renewcommand{\arraystretch}{1}
\scriptsize
%\footnotesize
\begin{tabular}{lrrrrlrrc||lrrrrlrr} 
\hline
PSR      & ObsDate &  \multicolumn{1}{c}{l}      & \multicolumn{1}{c}{b}        &Dist*    & DM*          &  RM   & $\sigma$      & &  PSR     &ObsDate  & \multicolumn{1}{c}{l}       &  \multicolumn{1}{c}{b}       &Dist*    & DM*          &  RM    & $\sigma$      \\  
         &   &\multicolumn{1}{c}{($^{\circ}$)}&\multicolumn{1}{c}{($^{\circ}$)} &(kpc)   &(cm$^{-3}$ pc)&\multicolumn{2}{c}{(rad~m$^{-2}$)} &    & &     &\multicolumn{1}{c}{($^{\circ}$)}&\multicolumn{1}{c}{($^{\circ}$)} &(kpc) &(cm$^{-3}$ pc)&\multicolumn{2}{c}{(rad~m$^{-2}$)} \\[1mm]
\hline                                                            
J0006+1834    &20210110  & 108.17   &      -42.99   &       0.9  &     11.41(55) &   -19.4 &   0.4 &   & J1745-0129    &20201129  &  23.84   &       14.03   &       5.5  &     89.3(3) &    36.3 &   0.9 \\     
J0011+08      &20210927  & 106.25   &      -53.45   &       5.5  &     25.6(9) &   -17.4 &   2.7 &     & J1745+1017  &20201115   &  34.87   &       19.25   &       1.2  &     23.9711(3) &    24.2 &   0.3  \\  
J0023+0923    &20201120  & 111.38   &      -52.85   &       1.8  &     14.3216(6) &    -5.5 &   3.0 &  & J1745+1252  &20210207   &  37.38   &       20.30   &      11.2  &     66.141(5) &    70.3 &   3.1  \\   
J0050+03      &20210117  & 122.49   &      -59.07   &      25.0  &     26.5(0) &   -25.6 &   2.2 &     & J1750+07    &20210208   &  32.84   &       16.93   &       3.8  &     55.6(20) &    54.9 &   2.7  \\    
J0058+4950    &20201120  & 124.05   &      -13.02   &       2.6  &     66.953(7) &  -114.2 &   0.5 &   & J1800-0125  &20201129   &  25.75   &       10.68   &       1.5  &     51.0(2) &    41.4 &   0.3  \\     
J0103+54      &20210927  & 124.67   &       -8.93   &       2.0  &     55.605(4) &   -74.5 &   5.4 &   & J1800+5034  &20201016   &  78.13   &       28.44   &       1.9  &     22.71(6) &    22.3 &   0.6  \\    
J0107+1322    &20210124  & 128.99   &      -49.31   &       2.3  &     21.671(1) &   -13.9 &   2.2 &   & J1802+0128  &20210117   &  28.61   &       11.56   &       7.7  &     97.8(4) &    18.4 &   1.1  \\     
J0122+1416    &20210110  & 134.03   &      -47.94   &       1.6  &     17.693(3) &   -14.8 &   4.9 &   &J1806+1023  &20210115    &  37.31   &       14.56   &       3.0  &     52.03(7) &    25.9 &   0.4  \\     
J0139+3336    &20210110  & 134.38   &      -28.17   &       1.5  &     21.23(1) &   -29.1 &   1.1 &    &J1806+2819  &20210110    &  54.65   &       21.67   &       1.3  &     18.6802(4) &    28.3 &   8.8  \\   
J0146+31      &20210927  & 136.69   &      -30.49   &       1.8  &     24.4(5) &   -32.0 &  26.6 &     &J1807+04    &20211123   &  31.46   &       11.66   &       2.4  &     53.0(3) &    92.6 &   1.3  \\      
J0229+20      &20210118  & 151.58   &      -36.42   &       2.1  &     28.5(5) &   -32.5 &   8.7 &     &J1807+0756  &20210117   &  35.15   &       13.28   &       7.9  &     89.29(3) &   148.1 &   1.9  \\     
J0241+16      &20210927  & 157.86   &      -39.24   &       0.9  &     19.3(8) &    -6.1 &   4.2 &     &J1809-0119  &20210118   &  26.96   &        8.63   &      11.1  &    138.3(2) &    21.7 &   0.3  \\      
J0244+14      &20210927  & 159.96   &      -40.02   &       2.2  &     29.3(7) &    -8.1 &   2.1 &     &J1809+17    &20211123   &  43.89   &       16.86   &       3.2  &     46.4(7) &    93.0 &   0.8  \\      
J0302+2252    &20201101  & 158.44   &      -30.82   &       1.0  &     18.9922(6) &    -7.8 &   0.4 &  &J1810+0705  &20210213   &  34.69   &       12.25   &       5.5  &     79.425(5) &   158.3 &   0.4  \\    
J0329+1654    &20201101  & 168.50   &      -31.68   &       2.6  &     40.821(36) &     2.6 &   4.1 &  &J1812+0226  &20201129   &  30.71   &        9.69   &       7.2  &    104.14(3) &   -22.7 &   0.5  \\     
J0337+1715    &20201114  & 169.99   &      -30.04   &       1.3  &     21.3162(3) &    26.4 &   0.1 &  &J1813+1822  &20210116   &  45.54   &       16.40   &       5.0  &     60.8(5) &   116.6 &   0.4  \\      
J0340+4130    &20201121  & 153.78   &      -11.02   &       1.6  &     49.5865(16) &    50.5 &   0.7 & &J1814+1130  &20210116   &  39.21   &       13.31   &       4.2  &     65(1) &    64.3 &   1.9  \\        
J0349+2340    &20210427  & 167.43   &      -23.38   &       3.7  &     62.962(5) &    96.6 &   3.9 &   &J1814+22    &20210207   &  49.49   &       17.78   &       6.5  &     62.313(11) &   118.5 &   2.4  \\   
J0358+4155    &20210108  & 156.11   &       -8.62   &       1.5  &     46.325(1) &    -7.0 &   0.2 &   &J1816+4510  &20201230   &  72.83   &       24.74   &       4.4  &     38.8874(4) &    37.6 &   4.1  \\   
J0453+1559    &20210108  & 184.13   &      -17.14   &       0.5  &     30.3053(3) &   -35.2 &   0.2 &  &J1821+4147  &20201031   &  69.54   &       22.91   &       4.4  &     40.673(3) &    35.9 &   0.4  \\    
J0457+23      &20210119  & 178.31   &      -12.01   &       1.5  &     59(0) &   -56.8 &   0.7 &       &J1822+1120  &20210116   &  39.87   &       11.57   &       7.8  &     95.2(6) &   108.9 &   1.2  \\      
J0518+5416    &20210110  & 155.92   &        9.56   &       1.4  &     42.330(5) &   -21.0 &  10.2 &   &J1828+0625  &20210118   &  36.07   &        8.02   &       1.0  &     22.4331(12) &    24.4 &   1.7  \\  
J0605+3757    &20210110  & 174.19   &        8.02   &       0.2  &     20.9462(4) &    -0.1 &   2.7 &  &J1829+2456  &20210115   &  53.34   &       15.61   &       0.9  &     13.700(5) &    -1.0 &   0.9  \\    
J0612+37216   &20210427  & 175.44   &        9.08   &       1.1  &     39.270(6) &    36.9 &   1.0 &   &J1832+27    &20211123   &  56.37   &       16.18   &       3.7  &     47.5(2) &    61.7 &   1.8  \\      
J0613+3731    &20210110  & 175.34   &        9.24   &       0.2  &     18.990(12) &    15.3 &   0.5 &  &J1834+10    &20210213   &  40.64   &        8.61   &       4.1  &     78.479(24) &    98.2 &   0.6  \\   
J0653+4706    &20210130  & 169.26   &       19.77   &       0.9  &     25.6(1) &    12.5 &   5.3 &     &J1838+1523  &20210117   &  45.35   &        9.69   &       3.5  &     68.26(3) &   160.4 &   0.4  \\     
J0711+0931    &20210111  & 206.67   &        8.78   &       1.2  &     46.0(8) &    60.9 &   0.2 &     &J1842+1332  &20210115   &  44.05   &        8.07   &       6.4  &    102.5(7) &   130.2 &   0.1  \\      
J0806+08      &20210928  & 214.02   &       20.34   &       1.9  &     48.4(13) &    10.2 &   5.0 &    &J1843+2024  &20210116   &  50.44   &       10.85   &       6.0  &     85.3(2) &   151.9 &   1.2  \\      
J0811+37      &20210928  & 183.63   &       31.25   &       0.4  &     15.9(4) &    -0.9 &   1.6 &     &J1849+2559  &20210427   &  56.24   &       11.87   &       6.1  &     75.0016(4) &    60.1 &   1.1  \\   
J0813+22      &20210928  & 200.91   &       27.48   &       2.4  &     52.29(5) &    14.0 &   6.6 &    &J1900+30    &20210211   &  61.76   &       11.80   &       6.8  &     71.8352(22) &   127.8 &   2.2  \\  
J0815+4611    &20201124  & 173.63   &       33.45   &       0.4  &     11.2738(3) &     2.3 &   0.4 &  &J1913+3732  &20201128   &  69.10   &       12.13   &       7.6  &     72.3623(41) &    -0.4 &   0.2  \\  
J0827+53      &20210928  & 165.36   &       35.74   &       1.6  &     23.11(2) &   -14.5 &   3.7 &    &J1916+3224  &20210212   &  64.63   &        9.43   &       7.9  &     84.105(2) &    55.6 &   0.8  \\    
J0848+16      &20210928  & 210.10   &       33.27   &       2.4  &     38.6(3) &    38.8 &   1.6 &     &J1923+4243  &20210106   &  74.72   &       12.64   &       4.7  &     52.99(5) &   -40.8 &   0.2  \\     
J0928+06      &20210928  & 226.82   &       37.56   &      25.0  &     49.8(5) &    11.0 &   6.0 &     &J1929+3817  &20201231   &  71.17   &        9.69   &       9.3  &     93.4(2) &   102.6 &   0.6  \\
J0944+4106    &20201124  & 180.44   &       49.38   &       2.7  &     21.41(3) &     1.3 &   0.8 &    &J1933+5335  &20210101   &  85.59   &       15.77   &       2.5  &     33.54(3) &     1.2 &   3.1  \\
J1017+3011    &20210124  & 198.43   &       56.27   &      25.0  &     27.150(2) &    19.2 &   0.8 &   &J1934+5219  &20201219   &  84.49   &       15.05   &       7.7  &     71.26(15) &     7.0 &   3.1  \\
J1038+0032    &20201120  & 247.15   &       48.47   &       5.9  &     26.340(12) &    14.3 &   5.2 &  &J1937-00    &20211123   &  37.82   &      -10.33   &       3.5  &     67.98(8) &   -15.4 &   0.8  \\
J1142+0119    &20210427  & 267.54   &       59.40   &       2.2  &     19.197(1) &    -5.2 &   1.7 &   &J1941+4320  &20201215   &  76.85   &        9.86   &       6.5  &     79.361(8) &    23.7 &   1.4  \\    
J1236-0159    &20201219  & 295.07   &       60.65   &       2.0  &     19.08(3) &     1.1 &   0.7 &    &J1947+0915  &20210118   &  47.71   &       -8.07   &       5.2  &     86.5(5) &  -114.7 &   0.2  \\      
J1312+0051    &20201215  & 314.84   &       63.23   &       1.5  &     15.345(1) &     4.0 &   0.6 &   &J1950+05    &20210213   &  44.86   &      -10.55   &       3.9  &     61.5(3) &   -59.3 &   0.7  \\      
J1312+1810    &20210118  & 332.95   &       79.76   &      20.6  &     25.0(1) &   -12.2 &   1.0 &     &J1953+1149  &20210116   &  50.71   &       -8.08   &      10.5  &    140.03(3) &   -46.1 &   3.8  \\     
J1501-0046    &20201218  & 356.58   &       48.05   &       2.1  &     22.2584(90) &    -1.9 &   1.0 & &J1954+1021  &20210117   &  49.52   &       -9.00   &       4.3  &     81.5(7) &   -18.6 &   0.3  \\      
J1518+0204A   &20210111  &   3.87   &       46.80   &       7.5  &     30.08(5) &     2.6 &   1.9 &    &J1954+4357  &20210109   &  78.53   &        8.17   &      10.1  &    130.30(5) &  -140.1 &   0.5  \\     
J1529+40      &20210122  &  66.18   &       54.91   &       0.5  &      6.61(16) &    -0.9 &   0.3 &   &J1957-0002  &20210427   &  40.63   &      -14.72   &       1.8  &     38.443(4) &    -8.1 &   2.6  \\    
J1538+2345    &20201216  &  37.32   &       52.39   &       1.3  &     14.909(1) &    10.6 &   0.2 &   &J2016+1948  &20201110   &  60.52   &       -8.68   &       2.2  &     33.8148(16) &  -123.2 &   0.2  \\  
J1544+4937    &20201219  &  79.17   &       50.17   &       3.0  &     23.2258(1) &    10.3 &   1.1 &  &J2017+0603  &20201108   &  48.62   &      -16.03   &       1.4  &     23.92344(9) &   -57.4 &   0.6  \\  
J1628+4406    &20201026  &  69.24   &       43.62   &       0.5  &      7.32981(2) &     1.6 &   0.4 & &J2027+2146  &20201111   &  63.54   &       -9.59   &      10.3  &     97.0915(48) &  -211.1 &   0.8  \\  
J1630+3734    &20201216  &  60.25   &       43.22   &       1.2  &     14.18009(7) &     1.1 &   0.2 & &J2033+1734  &20201111   &  60.86   &      -13.15   &       1.7  &     25.0864(12) &   -72.5 &   0.3  \\  
J1635+2332    &20201026  &  42.00   &       39.75   &      25.0  &     37.568(6) &    25.9 &   0.8 &   &J2040+1657  &20201110   &  61.29   &      -14.85   &       4.5  &     50.6919(14) &  -100.9 &   0.3  \\  
J1638+4005    &20210201  &  63.77   &       41.87   &      25.0  &     33.417(1) &    17.2 &   3.4 &   &J2042+0246  &20210111   &  48.99   &      -23.02   &       0.6  &      9.2694(2) &   -25.6 &   2.8  \\   
J1641+3627A   &20210115  &  59.00   &       40.91   &       7.1  &     30.4386(5) &    10.9 &   1.7 &  &J2048+2255  &20210111   &  67.45   &      -12.94   &       7.6  &     70.6847(22) &  -169.5 &   0.7  \\  
J1641+3627C   &20210115  &  59.00   &       40.91   &       7.1  &     30.1320(2) &     5.9 &   6.2 &  &J2051+1248  &20210110   &  59.36   &      -19.45   &       4.1  &     43.45(1) &   -68.6 &  11.5  \\     
J1641+3627D   &20210115  &  59.01   &       40.91   &       7.1  &     30.451(3) &    10.7 &   9.7 &   &J2105+07    &20210118   &  57.20   &      -25.05   &      25.0  &     52.6(0) &    10.6 &   0.5  \\      
J1643+1338    &20210203  &  31.26   &       34.36   &       4.6  &     35.821(1) &    36.0 &   1.0 &   &J2122+2426  &20210427   &  73.82   &      -17.93   &       0.6  &      8.500(5) &   -28.8 &   0.2  \\    
J1656+00      &20211028  &  19.20   &       25.47   &       3.9  &     46.9(0) &     5.1 &   3.5 &     &J2204+2700  &20210109   &  83.00   &      -22.65   &       3.2  &     34.8(11) &   -29.8 &   2.0  \\     
J1657+3304    &20210203  &  55.34   &       37.12   &       2.4  &     23.9746(6) &    22.2 &   1.3 &  &J2208+4056  &20201018   &  92.57   &      -12.11   &       0.8  &     11.837(9) &   -41.7 &   0.2  \\    
J1658+3630    &20210203  &  59.63   &       37.58   &       0.2  &      3.04387(3) &     8.2 &  31.5 & &J2209+22    &20210129   &  79.90   &      -27.79   &      25.0  &     45.4(6) &   -90.1 &   0.6  \\      
J1709+2313    &20201027  &  44.52   &       32.21   &       2.2  &     25.3474(2) &    39.1 &   0.8 &  &J2227+30    &20210126   &  89.66   &      -22.82   &       1.4  &     19.961(6) &   -60.7 &   2.5  \\    
J1710+4923    &20201013  &  75.93   &       36.45   &       0.5  &      7.08493(2) &     6.9 &   0.8 & &J2229+2643  &20201027   &  87.69   &      -26.28   &       1.8  &     22.7282(4) &   -60.0 &   0.7  \\   
J1715+46      &20210204  &  71.86   &       35.35   &       1.8  &     19.82(5) &    17.2 &   0.6 &    &J2234+0611  &20201020   &  72.99   &      -43.01   &       1.0  &     10.7670(2) &     5.6 &   3.1  \\   
J1722+35      &20211030  &  59.27   &       32.64   &       2.2  &     23.83(6) &    35.1 &   1.5 &    &J2234+0944  &20201021   &  76.28   &      -40.44   &       1.6  &     17.8323(2) &   -10.4 &   0.2  \\   
J1736+05      &20210206  &  29.59   &       19.21   &       2.5  &     42(8) &    53.9 &   1.1 &       &J2243+1518  &20201109   &  82.81   &      -37.38   &      25.0  &     39.9(2) &   -39.0 &   1.7  \\      
J1738+0333    &20201013  &  27.72   &       17.74   &       1.5  &     33.77312(4) &    33.2 &   0.3 & &J2302+4442  &20201031   & 103.40   &      -14.01   &       0.9  &     13.788120(0) &    17.7 &   0.2  \\ 
J1738+04      &20210206  &  28.39   &       18.21   &       1.1  &     23.1(6) &    26.7 &   2.8 &     &J2306+31    &20210126   &  97.96   &      -26.34   &      25.0  &     46.13(2) &   -79.0 &   1.1  \\     
J1739+0612    &20201013  &  30.26   &       18.86   &      25.0  &     95.4(1) &    26.6 &   0.3 &     &J2329+4743  &20201028   & 108.96   &      -12.91   &       2.4  &     44.012(2) &    -3.9 &   0.4  \\    
J1741+1351    &20201013  &  37.89   &       21.64   &       1.7  &     24.19871(14) &    63.2 &   3.6& &J2340+08    &20210125   &  94.82   &      -50.42   &       3.3  &     22.9(1) &    -4.7 &   2.5  \\      
J1743-0339    &20201114  &  21.65   &       13.40   &       0.2  &     30.26(11) &    48.2 &   2.8 &   &J2355+2246  &20201101   & 106.53   &      -38.32   &       2.2  &     23.1(7) &   -49.1 &   1.1  \\[1mm] 
\hline
\end{tabular}
\begin{tablenotes}
\item
*: The distance and DM value of these pulsars are obtained from the ATNF pulsar catalog \citep{mhth05}, and the distances are estimated by using YMW16 model from the DM values.
\end{tablenotes}
\end{table*}

\section{FAST polarization observations for pulsar rotation measures}
\label{obsdata}

We observe faint pulsars by using the FAST in the Galactic halo which previously had no measured RMs. The polarization observations were carried out in two projects [PID PT2020\_0164 and PID PT2021\_0051] with the L-band 19-beam receiver. The first was done between October 2020 to April 2021 towards pulsars with a good position. We track each pulsar by the central beam of the L-band 19-beam receiver for 10 minutes. The second project was carried out between September 2021 to November 2021 for pulsars with large position uncertainties by using the "SnapShotDec" observation mode. This "SnapShotDec" mode is modified from the "snapshot" mode designed for the FAST GPPS survey \citep{hww+21} but with the reference plane aligned with the Equatorial plane. One snapshot can have 4$\times$19 = 76 adjacent beams in a hexagonal sky region of 28$'$ wide in declination. The data of all 19 beams of 4 pointing are all recorded, and the signals of known pulsars are searched from all 76 beams, so that the position of a pulsar can be determined with an uncertainty of $1'$ \citep[see][]{hww+21}. The L-band 19-beam receiver is a dual-channel cryogenic system sensitive to orthogonal linear polarization with a system temperature of about 20 -- 25 K for different beams \citep{jyg+19}. The radio signals from the two polarization channels in the frequency range of 1000 to 1500~MHz are amplified, filtered, and then transferred to the digital room via optical fibers. The radio frequency signals are then sampled and channelized to 1024 or 2048 channels in the pulsar digital backend and then composed to 4 polarization for each channel \citep{hww+21}. These data are stored with a sampling time of 49.152 $\mu$s in search mode PSRFITS files \citep{hvm04}. For each pulsar, a 2-min observation of calibration signals of an amplitude of 10~K or 1~K switching on-off (1 second each) was made on the position, which will be used for system calibration.

Offline data analysis was performed using the PSRCHIVE software package \citep{hvm04}. The raw data were first de-dispersed based on the pulsar ephemeris obtained from the ATNF pulsar catalog \citep{mhth05} by using the DSPSR package \citep{vb11}. The radio frequency interference was examined, and corrupted frequency channels were excised using {\sc paz} and {\sc pazi}. The pulsar data were calibrated to compensate for instrumental gain and phase variation across the frequency band according to the calibration files obtained from the two minutes calibration on-off data, and then converted to Stokes parameters. More details and reliability of polarimetric calibration for each beam of the FAST 19-beam receiver can be found in \citet{whj+22}.

\begin{table}[H]
\centering 
\small 
\caption{Comparison of our FAST RMs with published values}
\label{RMcomp}
\renewcommand{\arraystretch}{0.8}
%\scriptsize
\begin{tabular}{lrrl} 
\hline
PSR         &  FAST RM             & RM1 or RM2     & Ref   \\    
name        &  (rad m$^{-2}$)  & (rad m$^{-2}$)    &      \\
\hline                                                            
J0006+1834  &  -19.4 $\pm$ 0.4 &  -20 $\pm$ 3    & [1]   \\
J0023+0923  &  -5.5  $\pm$ 3.0 & -5.1 $\pm$ 0.9  & [2]*  \\
J0340+4130  &   50.5 $\pm$ 0.7 & 56.1 $\pm$ 0.7  & [1]  \\
            &                  & 53.5 $\pm$ 1.1  & [3]  \\
J0453+1559  &  -35.2 $\pm$ 0.2 &  -35 $\pm$ 2    & [1]  \\
            &                  &-35.3 $\pm$ 0.3  & [2]  \\
J0613+3731  &   15.3 $\pm$ 0.5 &   16 $\pm$ 2    & [1]   \\
J0711+0931  &   60.9 $\pm$ 0.2 & 62.8 $\pm$ 1.1  & [1]   \\
J1038+0032  &   14.3 $\pm$ 5.2 &   20 $\pm$ 5    & [2]*  \\
J1142+0119  &   -5.2 $\pm$ 1.7 & 0.31 $\pm$ 0.11 & [2]*  \\
J1312+0051  &    4.0 $\pm$ 0.6 &    2 $\pm$ 1    & [2]   \\
J1538+2345  &   10.6 $\pm$ 0.2 & 11.5 $\pm$ 1.1  & [1]   \\
J1544+4937  &   10.3 $\pm$ 1.1 &  9.8 $\pm$ 1.9  & [1]   \\
J1641+3627A &   10.9 $\pm$ 1.7 &   13 $\pm$ 3    & [1]   \\
J1709+2313  &   39.1 $\pm$ 0.8 &   37 $\pm$ 2    & [2]   \\
J1710+4923  &    6.9 $\pm$ 0.8 &   12 $\pm$ 2    & [1]   \\
J1736+05    &   53.9 $\pm$ 1.1 &   44 $\pm$ 3    & [1]   \\
J1738+0333  &   33.2 $\pm$ 0.3 &   33 $\pm$ 3    & [1]  \\
            &                  &  34.7 $\pm$ 0.5 & [2]  \\
J1741+1351  &   63.2 $\pm$ 3.6 & 63.5 $\pm$ 0.7  & [2]*  \\
J1745+1017  &   24.2 $\pm$ 0.3 & 27.2 $\pm$ 0.3  & [2]  \\
J1828+0625  &   24.4 $\pm$ 1.7 &   20 $\pm$  4   & [2]  \\
J1829+2456  &   -1.0 $\pm$ 0.9 &  0.3 $\pm$ 0.6  & [2]*  \\
J1834+10    &   98.2 $\pm$ 0.6 & 97.0 $\pm$ 1.7  & [1]  \\
J1900+30    &  127.8 $\pm$ 2.2 &  121 $\pm$ 2    & [1]*  \\
J2016+1948  & -123.2 $\pm$ 0.2 & -121 $\pm$ 4    & [1]  \\
J2017+0603  &  -57.4 $\pm$ 0.6 &  -59 $\pm$ 3    & [2]  \\
J2027+2146  & -211.1 $\pm$ 0.8 & -210 $\pm$ 2    & [1]  \\
J2033+1734  &  -72.5 $\pm$ 0.3 &-71.5 $\pm$ 0.3  & [2]  \\
J2040+1657  & -100.9 $\pm$ 0.3 &  -98 $\pm$ 2    & [1]  \\
J2042+0246  &  -25.6 $\pm$ 2.8 &  -21 $\pm$ 2    & [2]*  \\
J2208+4056  &  -41.7 $\pm$ 0.2 &  -40 $\pm$ 3    & [1]  \\
J2227+30    &  -60.7 $\pm$ 2.5 &-58.3 $\pm$ 1.9  & [1]*  \\
J2229+2643  &  -60.0 $\pm$ 0.7 &-61.2 $\pm$ 0.3  & [2]* \\
J2234+0611  &    5.6 $\pm$ 3.1 &    4 $\pm$ 1    & [2]*  \\
J2234+0944  &  -10.4 $\pm$ 0.2 &-11.5 $\pm$ 0.5  & [2]  \\
J2243+1518  &  -39.0 $\pm$ 1.7 &-35.5 $\pm$ 0.5  & [1]*  \\
J2302+4442  &   17.7 $\pm$ 0.2 & 19.1 $\pm$ 1.6  & [1]   \\
            &                  & 18.4 $\pm$ 0.4  & [3]   \\
J2340+08    &   -4.7 $\pm$ 2.5 &   -7 $\pm$ 2    & [1]*  \\
\hline
\end{tabular}
\begin{tablenotes}
\item  References. [1]: \citet{npn+20}; [2]: \citet{sbm+22}; [3]: \citet{wmg+22}; "*" denotes the source for the best value (not from this work).
\end{tablenotes}
\end{table}

For the pulsars with large position uncertainties, pulsar search was made first for every beam in a cover of snapshot observation following \citet{hww+21}, and then the polarization data of the detection beam are analysed with the same procedures as above. We determined the positions for 12 pulsars, often very bright for FAST, by such beam pointing following the procedure described by \citet{hww+21}. Signal-to-noise ratios of pulsar profiles can then be computed for several nearby beams so that the position can be determined according to the positions of beam centers and the signal-to-noise ratios of these nearby beams. The accuracy of the determined position is better than $1'$. Our observations also find that the previous period of four pulsars J0711+0931, J0848+16, J1750+07 and J1802+0128 are just harmonics. Their corrected parameters are listed in Table~\ref{parupdate}.

The rotation measure from the resulting polarization profiles was determined by using {\sc rmfit}\footnote{http://psrchive.sourceforge.net/manuals/rmfit}, following \citet{hmvd18}. First, an initial guess of RM was found by searching for peak linear polarization in the RM range of $\pm1000$~rad~m$^{-2}$. The total linear polarization $L=(Q^2+U^2)^{1/2}$ was computed by summing across all phase bins and the frequency band after correction of Faraday rotation at each trial RM. Second, the first guess value was iteratively corrected by using the RM-refinement algorithm: integrating the data separately in the two halves of the band by taking the current RM and then obtaining the correction to the RM from computing the weighted differential polarization angle. The uncertainty of RM is calculated from the weighted average uncertainty of position angles in the two halves of the band. Only pulse phase bins with linearly polarized flux which is 3-sigma above the off-pulse in both halves are included in the estimate of differential position angles. The final RM value was then derived by subtracting the ionospheric RM calculated with IonFR\footnote{https://sourceforge.net/projects/ionfarrot} \citep{ssh+13}. The error in the ionospheric RM was incorporated into the final error of RM value.

We determined rotation measures for 118 pulsars in the first FAST observation project and 16 pulsars in the second, as listed in Table~\ref{RMtable}. All polarization profiles are published in \citet{whj+22} as a part of the FAST pulsar database. Figure~\ref{psrsky} shows the sky distributions of all halo pulsars together with disk pulsars in the Galactic coordinates. Of the total 134 pulsars, RMs of 36 pulsars were also measured recently by \citet{npn+20}, \citet{sbm+22} and \citet{wmg+22}, consistent with our measurements with a smaller uncertainty as listed in Table~\ref{RMcomp}. Marginally significant differences ($>3\sigma$) are only found for PSRs J0340+4130 ($5.6\sigma$), J1142+0119 (3.2$\sigma$),  J1736+05 (3.1$\sigma$), and J1745+1017 (7.1$\sigma$), probably due to slightly underestimated uncertainties of some measurements or ionospheric RM corrections.

\begin{figure*}[!t]
  \centering 
  \includegraphics[width=0.46\textwidth]{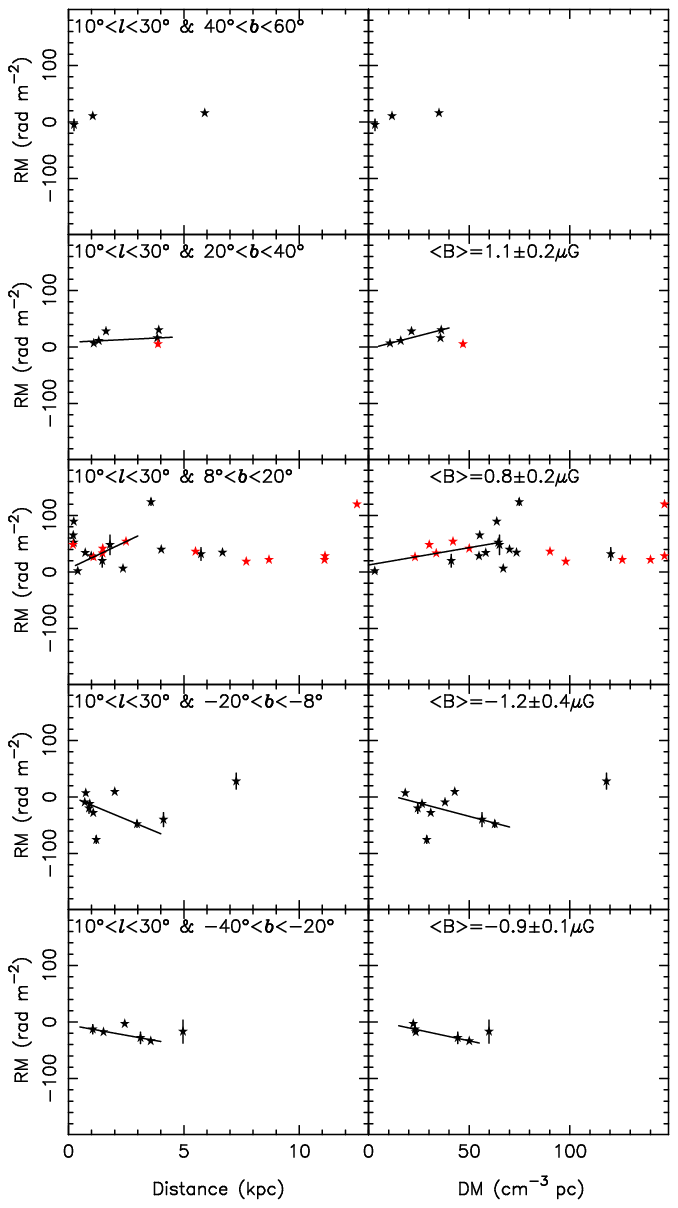} \hspace{1mm}
  \includegraphics[width=0.46\textwidth]{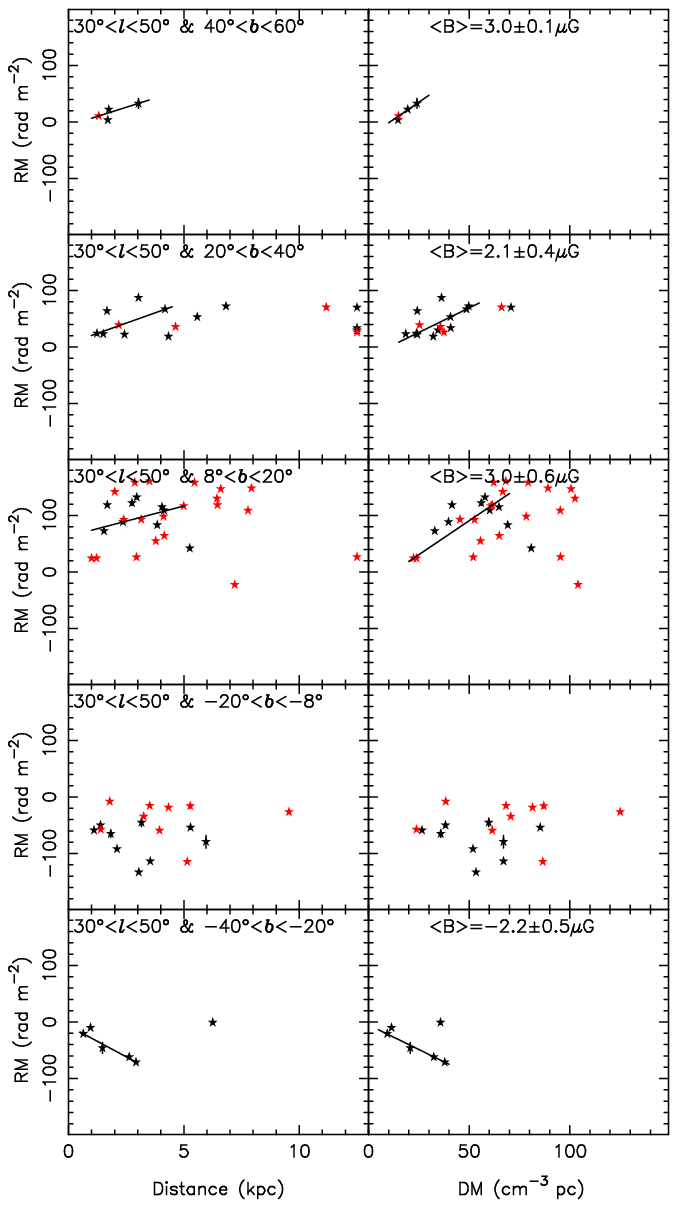}
  \caption{RMs (with error-bar) versus distance and DM for pulsars in the binned halo regions of the Galactic latitude of $8^{\circ}<|b|<60^{\circ}$ and in the Galactic longitude ranges of $10^{\circ}<l<30^{\circ}$ ({\it left}) and $30^{\circ}<l<50^{\circ}$ ({\it right}). The red stars denote the newly obtained RMs for faint pulsars by FAST. The black lines represent the robust straight-line fitting \citep[see \S15.7 in][]{ptvf92} to RMs for pulsars. The mean line-of-sight magnetic field and its statistical uncertainty are given inside the panels, with positive slopes corresponding to mean fields towards us. The uncertainties of DMs are typically less than the symbol size and not plotted. The distance uncertainties for two thirds of pulsars are smaller than 20\% \citep{ymw17}.}
\label{strenin1}
\end{figure*}

\begin{figure*}[!t]
  \centering 
  \includegraphics[width=0.46\textwidth]{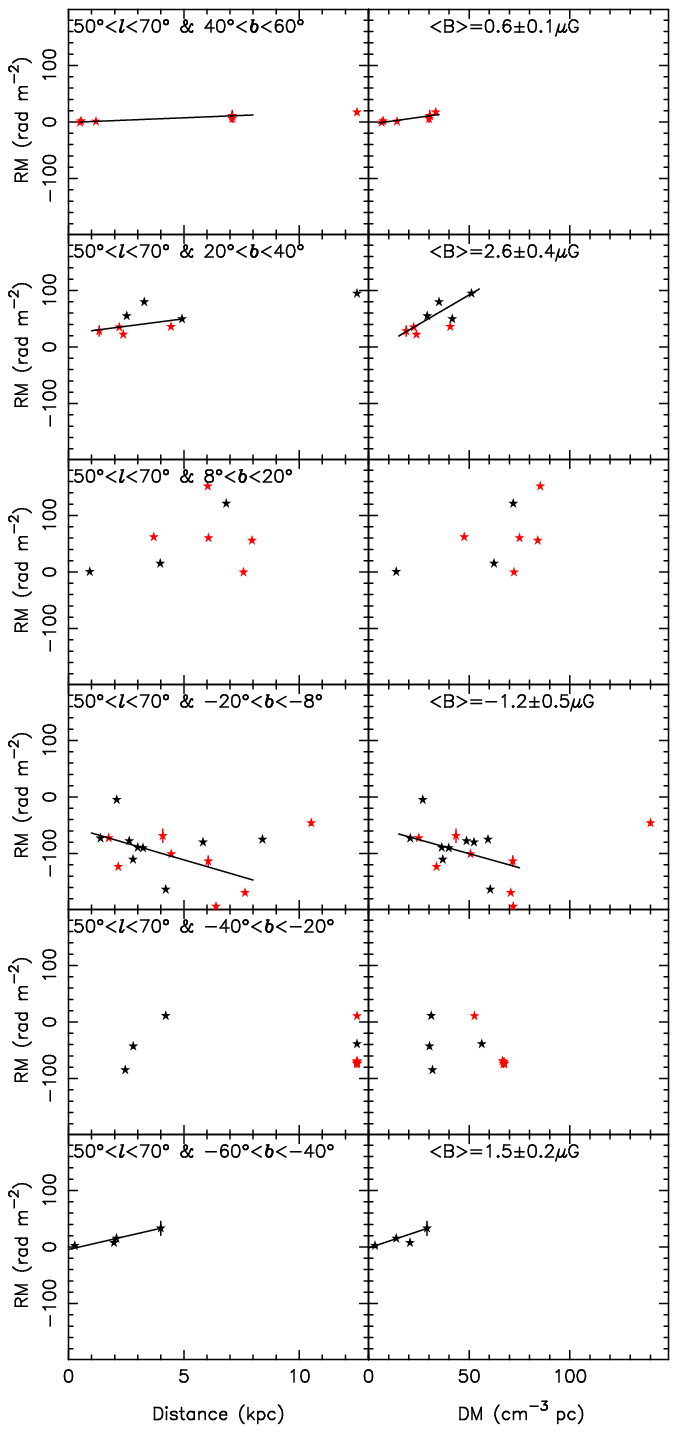} \hspace{1mm}
  \includegraphics[width=0.46\textwidth]{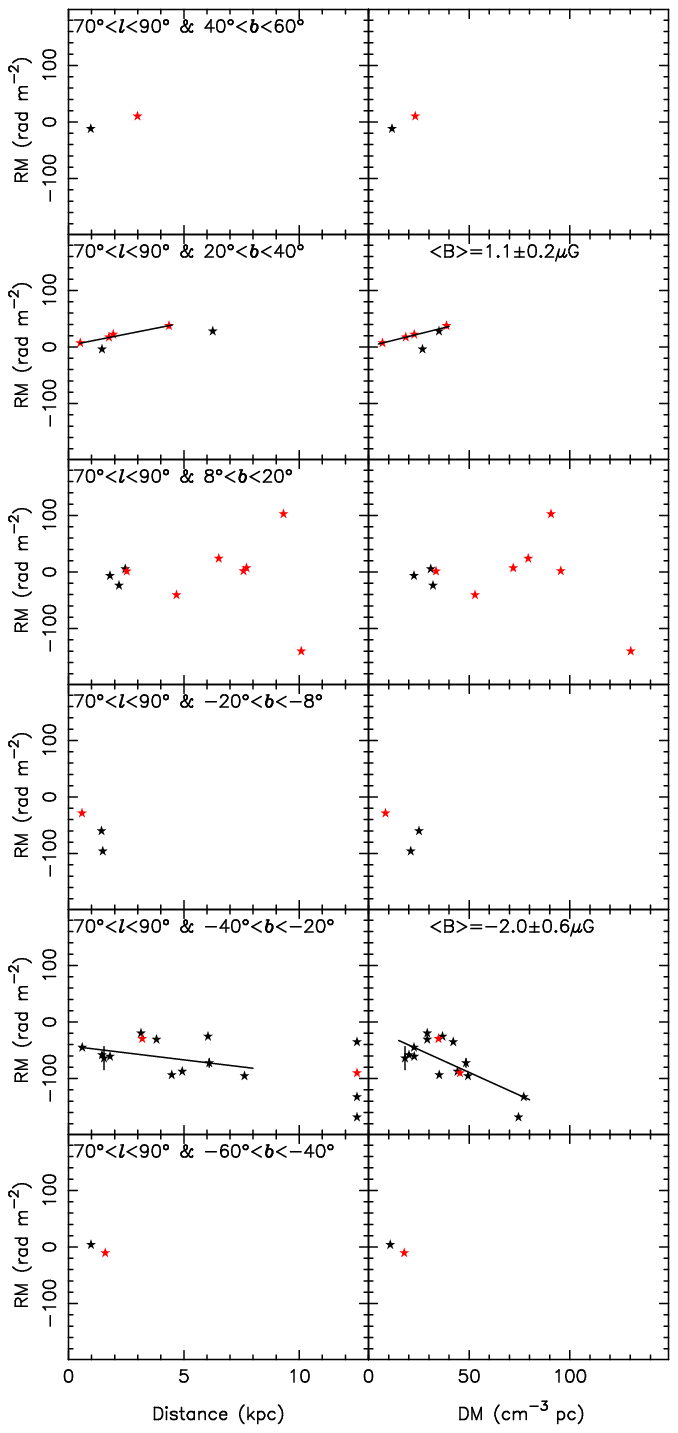}
  \caption{Same as Figure~\ref{strenin1} but for RMs for pulsars in the binned halo regions of the Galactic longitude ranges $50^{\circ}<l<70^{\circ}$ ({\it left}) and $70^{\circ}<l<90^{\circ}$ ({\it right}).}
\label{strenin2}
\end{figure*}

\section{Magnetic fields in the Galactic halo}
\label{halofield}

Though the magnetic field structures in the Galactic halo have been qualitatively proposed by \citet{hmbb97,hmq99}, and then quantitatively modelled by, e.g.,  \citet{ps03,srwe08,ptkn11,jf12,tf17,xh19}, many key parameters such as the scale radius, the scale height, the field strength and its variations as a function of radius and height, have not been well determined yet. There have been some efforts to analysis of pulsar RMs for the scale height of the Galactic halo fields \citep{sbg+19} and the field strength \citep{xh19}. Based on much more pulsar data, we can investigate the nature of magnetic fields in the halo as follows.

\subsection{The strength of magnetic fields in the Galactic halo}
\label{halostrength31}

Since the RM distributions of pulsars and EGRS shows coherent structures on scales of several degrees to tens of degrees on the sky, if the halo fields are coherent on the large scales, the strength of magnetic fields can be estimated from the variations of pulsar RMs against DMs in a small sky area where pulsars are located in similar directions.  

Considering the number density of pulsars, we divide the middle Galactic latitude region into a number of modest-sized bins to examine the RM trends over different distance intervals to pulsars. As shown in  Figure~\ref{psrsky}, we make three Galactic latitude ranges for each hemisphere above and below the Galactic plane: $|b|$ = ($8^{\circ}, 20^{\circ}$), ($20^{\circ}, 40^{\circ}$) and ($40^{\circ}, 60^{\circ}$). Then each Galactic latitude range is further divided into six longitude ranges, four in the first quadrant from $l=10^{\circ}$ to $l=90^{\circ}$ with a width of $20^{\circ}$ and two in the second quadrant from $l=90^{\circ}$ to $l=180^{\circ}$ with a width of $45^{\circ}$. The Galactic longitude ranges $l<10^{\circ}$ excluded because the large-scale azimuthal fields become nearly perpendicular to the lines of sight close to the meridian through the Galactic Centre, and hence not easy to measure using Faraday rotation.

\begin{figure}[H]
  \centering \includegraphics[width=0.42\textwidth]{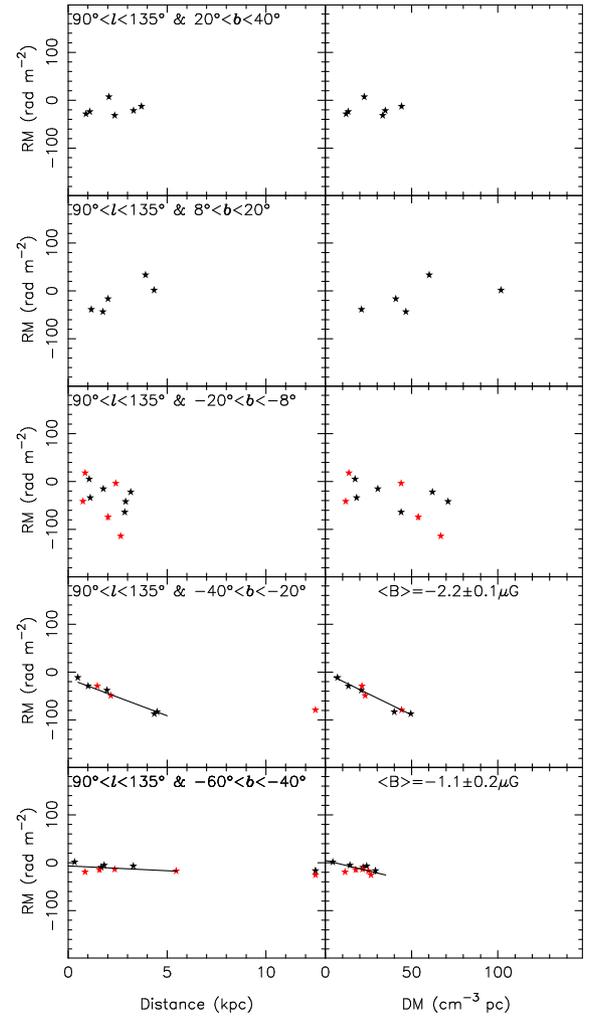} %\hspace{1mm}
  \caption{Same as Figure~\ref{strenin1} but for RMs for pulsars
  in the binned halo regions of the Galactic longitude range $90^{\circ}<l<135^{\circ}$.}
\label{strenout}
\end{figure}

Since distances of individual pulsars determined from the Galactic electron  density distribution models have unpredictable uncertainties and RMs of individual pulsars are subjected to small-scale fluctuations of the magnetic fields, following \citet{hml+06,hmvd18}, we quantify the RM trends by fitting the variations of RM against DM and distance for groups of pulsars over some distance intervals where a significant correlation between RM and DM exists. To determine the RM--DM slope, i.e., the averaged field strength along the line of sight, the Maximum Likelihood Robust Estimate routine \citep[see \S15.7 in][]{ptvf92} is used (i.e., the {\sc medfit} subroutine). The ``robust'' fitting is necessary to minimize the effects of outliers that could influence the slope of the fitted line resulting from un-modeled fluctuations of electron density and/or magnetic field along the path. A straight line is plotted in Figure~\ref{strenin1}, \ref{strenin2} and \ref{strenout} for RMs over distance/DM ranges where the result is significant at least above 2-sigma of none-zero\footnote{The straight line fitting over the DM range reflects a general trend which indicates the mean magnetic field. The scatter around the fitted line is mainly caused by the random magnetic field, rather than small measurement uncertainties.  Therefore, the returned uncertainty of the fit represents the random magnetic field. Therefore, the 3-sigma criterion should not be a compulsory criterion if the random magnetic field is comparable to the large-scale mean magnetic field. For example, 0.5$\pm$0.2~$\mu$G is a good fitting, not necessary over $3\sigma$.} The averaging over groups of pulsars will minimize the effects of distance uncertainties of individual pulsars. The uncertainty of the slope is taken as the mean absolute deviation of RMs from the fitted line divided by the DM range for the fitting. The scatter of data points around the fitted line reflects the random component of magnetic fields on the line of sight rather than the effects of small RM measurement uncertainties. Positive slopes of RM--DM denote magnetic fields with a direction towards us, while negative slopes denote fields pointing away from us. To improve the field estimates reliability, we have omitted the pulsar RMs with  uncertainties larger than 30 rad~m$^{-2}$.

We emphasize that the estimated magnetic fields are derived merely from the fits to RM versus DM. Figure~\ref{strenin1} and \ref{strenin2} show pulsar RMs versus distance and DM in the halo areas towards the inner Galaxy, if there are more than 4 pulsars in the binned region, and Figure~\ref{strenout} show these relations towards the outer Galaxy. From the top sub-panels down, each shows RM relations in the regions from the northern hemisphere to the southern hemisphere. All derived values of magnetic field strength have more than or about 3-sigma significance of none-zero as listed in Table~\ref{Bfield}.

Towards the inner halo regions, as shown in Figure~\ref{strenin1} and \ref{strenin2} and values for the longitude ranges $10^{\circ}<l<90^{\circ}$ in Table~\ref{Bfield}, the estimated magnetic fields have positive values in the northern hemisphere and negative in the southern hemisphere, consistent with a bi-toroidal configuration of opposite field directions above and below the Galactic plane proposed by \citet{hmbb97,hmq99}. 
The field strengths above and below the Galactic plane are roughly consistent with each other within the uncertainties, around 2~$\mu$G at $30^{\circ}<l<70^{\circ}$.
In most of the regions where the pulsar RM-DM data can be fitted (at least 2-sigma significance of non-zero), the far end of distance range for the fit is around 4--6~kpc. The RM data of pulsars have a larger dispersion at larger distances. 
The RM-DM data in some regions cannot be fitted due to either a small number of pulsars or a large scatter of RM values, but the RM tendency indicates a field as same as neighboring regions in the same hemisphere.

Towards the outer Galactic halo, as shown in Figure~\ref{strenout}, good RM-DM relations only exist in the southern hemisphere at $90^{\circ}<l<135^{\circ}$. The maximum strength of line-of-sight component  is up to around 2.2~$\mu$G, which indicates strong large-scale toroidal fields also exist in the south outer halo.

\subsection{The scale height of the halo magnetic field}

Early estimates of the scale height of the Galactic magnetic field are mainly based on RMs of EGRS. By assuming a uniform magnetic field model, \citet{sk80} estimate the full thickness of the magnetic disk of around 1.4~kpc by fitting the RM variance of EGRS in bins of 10$^{\circ}$ in galactic latitude. \citet{hq94} combine the uniform model of \citet{sk80} with variations of both field strength and electron density to estimate the scale height of about 1.2$\pm$0.4 kpc. The scale height would be much larger if a smaller electron density is adopted \citep{gmcm08}.

\begin{table}[H]
\centering
\footnotesize
\caption{Magnetic fields in the Galactic halo regions derived from pulsar DMs and RMs}
\label{Bfield}
\setlength{\tabcolsep}{5pt}
\renewcommand{\arraystretch}{1}
%\scriptsize
\begin{tabular}{cccccc} 
\hline
  b-range             &  D-range     &  DM-range      &   No.    & B$_{||}$      & B-field      \\
                      & (kpc)        & (cm$^{-3}$~pc)   &   PSRs      & ($\mu$G)       & Direction \\
\hline
\multicolumn{6}{c}{$10^{\circ}<l<30^{\circ}$}\\
\hline                                                            
($20^{\circ},40^{\circ}$)   &0.5--4.5      &5--40           &   5         &+1.1$\pm$0.2   & ccw       \\
($8^{\circ},20^{\circ}$)    &0.3--3.0      &0--65           &   12        &+0.8$\pm$0.2   & ccw       \\
($-20^{\circ},-8^{\circ}$)  &0.5--4.0      & 15--70         &   9         &-1.2$\pm$0.4   & cw        \\
($-40^{\circ},-20^{\circ}$) &0.5--4.0      & 15--55         &   5         &-0.9$\pm$0.1   & cw        \\
\hline
\multicolumn{6}{c}{$30^{\circ}<l<50^{\circ}$}\\
\hline                                                            
($40^{\circ},60^{\circ}$)   &1.0--3.5      &10--30          &   4         &+3.0$\pm$0.1   & ccw       \\
($20^{\circ},40^{\circ}$)   &1.0--4.5      &15--55          &   14        &+2.1$\pm$0.4   & ccw       \\
($8^{\circ},20^{\circ}$)    &1.0--5.0      &20--70          &   20        &+3.0$\pm$0.6   & ccw       \\
($-40^{\circ},-20^{\circ}$) &0.5--3.0      & 5--40          &   6         &-2.2$\pm$0.5   & cw        \\
\hline
\multicolumn{6}{c}{$50^{\circ}<l<70^{\circ}$}\\
\hline                                                            
($40^{\circ},60^{\circ}$)   &0.5--8        &5--35           &   7         &+0.6$\pm$0.1   & ccw       \\
($20^{\circ},40^{\circ}$)   &1--5          &15--55          &   8         &+2.6$\pm$0.4   & ccw       \\
($-20^{\circ},-8^{\circ}$)  &1--8          &15--75          &   16        &-1.2$\pm$0.5   & cw        \\
($-60^{\circ},-40^{\circ}$) & 0--4         & 0--30           &   4       & +1.5$\pm$0.2   &  ccw      \\
\hline
\multicolumn{6}{c}{$70^{\circ}<l<90^{\circ}$}\\
\hline                                                            
($20^{\circ},40^{\circ}$)   &0.5--4.5      &5--40           &   6         &+1.1$\pm$0.2   & ccw       \\
($-40^{\circ},-20^{\circ}$) &0.5--8        &15--80          &   17        &-2.0$\pm$0.6   & cw        \\
\hline
\multicolumn{6}{c}{$90^{\circ}<l<135^{\circ}$}\\
\hline                                                            

($-40^{\circ},-20^{\circ}$) &0.5--5      &5--50           &   8         &-2.2$\pm$0.1   & cw        \\
($-60^{\circ},-40^{\circ}$) &0--5.5        &0--35           &   10        &-1.1$\pm$0.2   & cw        \\
\hline
\multicolumn{6}{l}{cw means clockwise, and ccw means counterclockwise.} 
\end{tabular}
\end{table}

Direct estimation of the parallel magnetic field along the sight lines decouples the scale heights for electron density and magnetic field.  In comparison with the recent LOFAR survey \citep{sbg+19} and CHIME survey \citep{npn+20}, our new measurements have largely increased the number of RMs for the distant pulsars in the first and second Galactic quadrant (see Figure~\ref{psrsky}). With the Z distribution of halo pulsars in a large range, the scale height of magnetic fields in the Galactic halo can be better constrained.

\begin{figure}[H]
\centering \includegraphics[angle=0,width=0.44\textwidth]{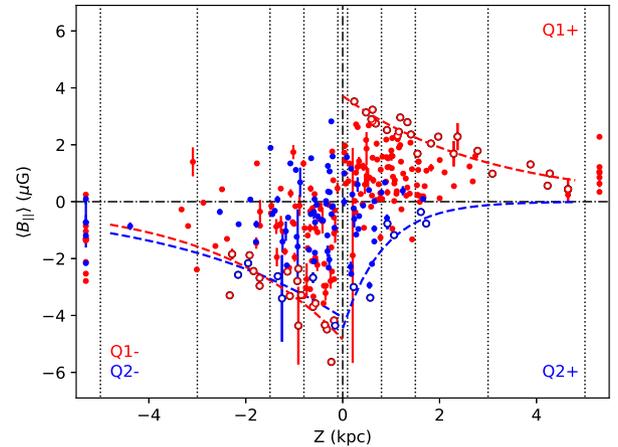}
\caption{The parallel magnetic fields along the sight-line path as a function of vertical height above/below the Galactic plane for the halo pulsars of $|b|>8^{\circ}$ located in the first Galactic quadrant (Q1: $10^{\circ}<l<90^{\circ}$) and the second quadrant (Q2: $90^{\circ}<l<180^{\circ}$). The vertical dotted lines indicate the height at $|Z|=$ 0.1, 0.8, 1.5, 3.0 and 5.0 kpc. Pulsars with heights beyond 6 kpc are plotted at $|Z|=$ 5.3~kpc. Red points represent pulsars in Q1, and blue points in Q2. The "outermost" points in the four areas labeled Q1+, Q1-, Q2+ and Q2- are indicated by open circles. Dashed lines outline fitting results to the "outermost" data points.}
\label{scalehightb8}
\end{figure}

Figure~\ref{scalehightb8} shows the averaged parallel magnetic field $\left < B_{||} \right >$ along the line of sight derived by using Equation (\ref{meanB1}) towards halo pulsars as a function of their vertical heights from the Galactic plane. Again, pulsars with RM uncertainties larger than 30 rad m$^{-2}$ are discarded. Pulsars in Q1 indicate positive (negative) magnetic field values above (below) the Galactic plane, and most of those in Q2 have negative values. According to the data number distribution, we divide the data points into four Z-height ranges, $|Z|=$ (0.1, 0.8), (0.8, 1.5), (1.5, 3.0) and (3.0, 5.0) kpc. Referring to \citet{sbg+19}, the envelopes of $\left<B_{||}\right>$ distribution can be characterized by the "outermost" points, identified as the largest absolute values in the ranges. Exponential envelopes are clearly seen for magnetic fields at both positive and negative heights, as expressed by  
\begin{equation}
  \left<B_{||}\right> = \left<B_{||}\right>_{0} exp(-|Z|/H), 
\end{equation}
where $\left<B_{||}\right>_{0}$ is the constant field value at $Z=0$ and $H$ is 
the scale height of magnetic fields. 
The magnetic scale heights derived in the areas of Q1+ and Q1-  have a value of 2.7$\sim$3.0~kpc. The scale height in the Q2- area has a value of 3.7~kpc but with large uncertainty. In contrast, the Q2+ area has a much smaller scale height, only 0.8~kpc. 
The large scale heights in the three areas of Q1+, Q1-, and Q2- are also supported by distant pulsars with vertical height beyond 6~kpc, as seen in the figure, though distances of those distant pulsars are likely overestimated. The fits to all pulsar data for the three areas of Q1\&Q2- gives a value of 2.7$\pm$0.3 kpc, slightly larger than the scale height of 2.0~kpc estimated by \citet{sbg+19}. Note that this derived value of 2.7~kpc is the lower limit on the true scale height of magnetic fields.

\begin{figure*}[!t]
\centering \includegraphics[angle=-90,width=0.7\textwidth]{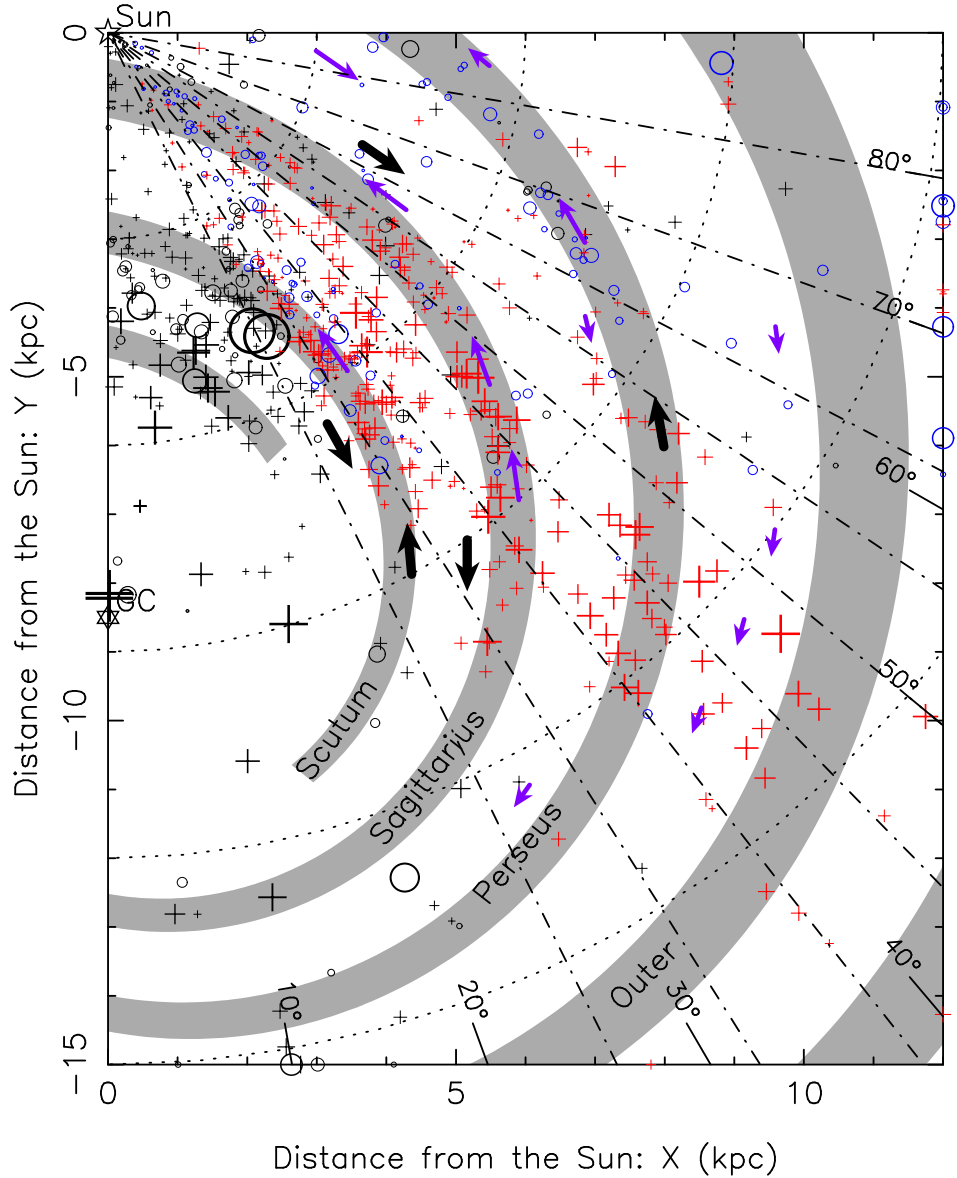}
\caption{RM distribution of pulsars of $|b|<8^{\circ}$ in the first Galactic quadrant. The symbols are as same as in Figure~\ref{psrsky}. New measurements are indicated by red crosses and blue circles for positive and negative values. The approximate locations of spiral arms \citep[see][]{hh14} are indicated in shadow. The dot-dashed lines give the longitude boundaries for the RM analyses in this paper. Distances to pulsars are estimated by using the YMW16 Galactic electron density distribution model \citep{ymw17}. Two thirds of them have an uncertainty less than 20\%. Pulsars with distances too large to show in the plot ranges are plotted at the boundaries. Magnetic field directions inferred from RM-DM fits (at least 2-sigma significance of non-zero) of pulsars are plotted with thick long arrows representing new determinations and with thin arrows for previously known directions. Short arrows give a inferred direction derived from the RM comparison of EGRS and most distant pulsars.}
\label{mfdirection}
\end{figure*}

\begin{figure*}%[!t]
\centering \includegraphics[width=0.62\textwidth]{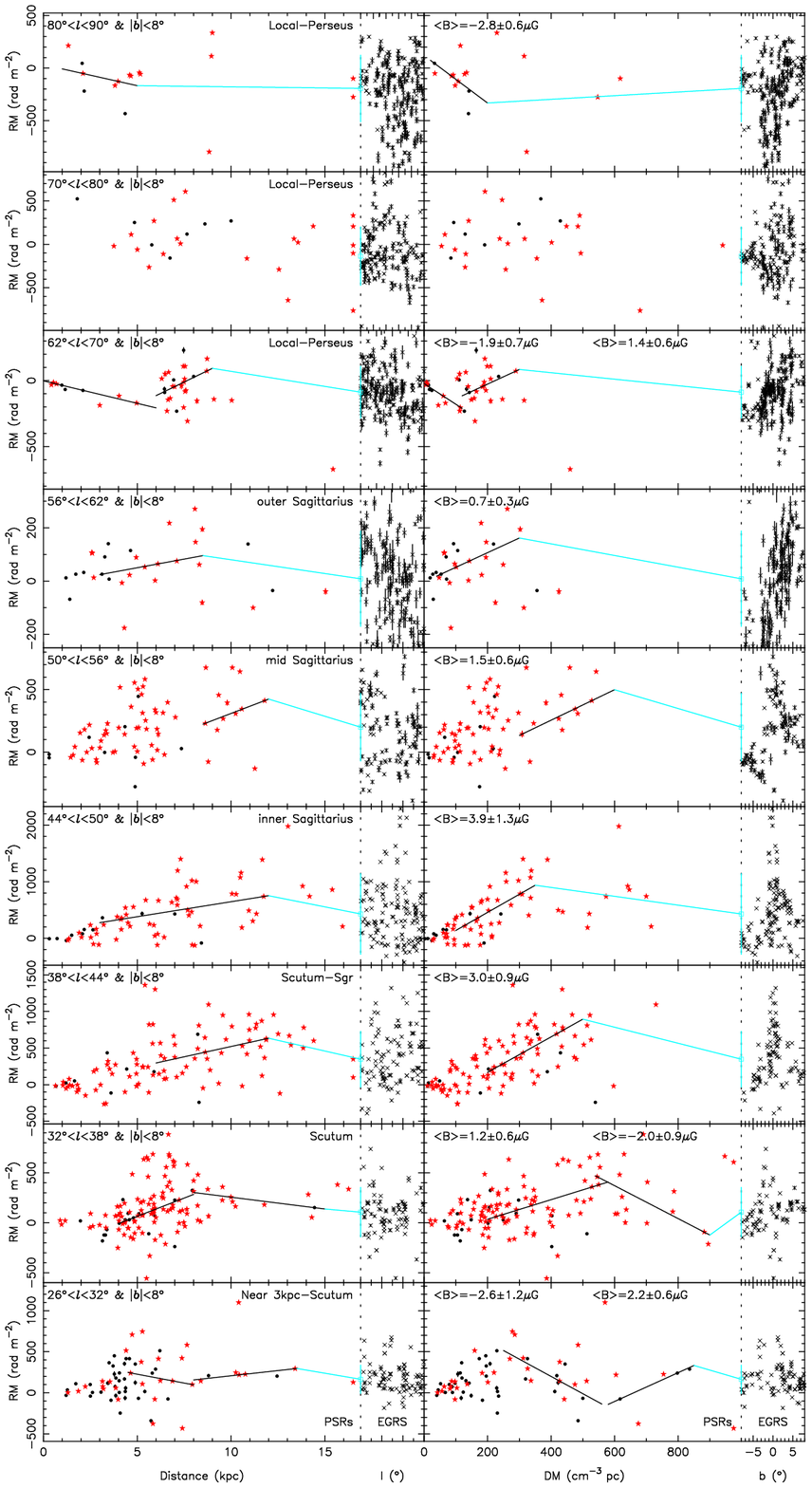}
\caption{RMs for pulsars versus distance and DM, and RMs for EGRS versus the Galactic longitude (scale ticks are given but not labeled) and the Galactic latitude in several longitude intervals at $26^{\circ}<l<90^{\circ}$ of the Galactic disk. The red stars denote the newly obtained RMs by the FAST, and the black dots represent previously known RMs. The black lines represent the robust straight-line fitting \citep[see \S15.7 in][]{ptvf92} to RMs for pulsars.  The blue lines indicate the comparison of RMs of the most distant pulsars to the median of the EGRS RMs that are denoted by the blue squares with error bars representing the standard deviation from the median in the region. The EGRS RMs are collected from the literature \citep{lsc90,ccsk92,btj03,tss09,vbs+11,sls+19,scw+19,mmob20,vbo+21}.}
\label{rmdm4q1}
\end{figure*}

\section{Magnetic fields in the first quadrant of the Galactic disk}
\label{diskfield}

The GPPS survey has discovered more than 500 new pulsars in the Galactic disk of $|b|<10\degree$, and redetected most of previously known pulsars in the survey region \citep{hww+21}. Polarization data are recorded for most of the pulsars during the survey observation. In the FAST pulsar database of \citet{whj+22}, we determined RMs for the newly discovered pulsars and many known pulsars (including 134 RMs of halo pulsars in this work). 
Among them, RMs of 311 pulsars are first time measured, and RMs of 161 pulsars are updated. 
This large number of RMs for weak distant pulsars in the Galactic longitude of  $26^{\circ}<l<90^{\circ}$ in the first quadrant (see Figure~\ref{mfdirection}) increased by a factor of more than two compared to the previous work in the longitude ranges \citep{hmvd18}. This has enabled us to explore the magnetic fields in farther arms up to 15~kpc in the first Galactic quadrant.

\begin{table*}[!t]
\centering
\small 
\caption{Magnetic fields in the first Galactic quadrant}
\label{BdiskQ1}
\renewcommand{\arraystretch}{0.8}
%\scriptsize
\begin{tabular}{llllcclll} 
\hline
Region                &l-range &  D-range     &  DM-range      &   No. PSRs    & B$_{||}$      & B-field           &  Arrow l             & Arrow D   \\
                      &($^{\circ}$) & (kpc)        & (cm$^{-3}$~pc)   &  or EGRS      & ($\mu$G)       & Direction       &($^{\circ}$)      & (kpc) \\
\hline                                                            
Near 3-kpc--Scutum\_1   & $26^{\circ}<l<32^{\circ}$   & 4.5--8.0          &  250--560     &   21       &  -2.6$\pm$1.2          &  cw   & 29     & 6.5 \\
Near 3-kpc--Scutum\_2   & $26^{\circ}<l<32^{\circ}$   & 8.0--13.5         &  580--850     &   6       &  2.2$\pm$0.6          &  ccw   & 29     & 9.0 \\
Near 3-kpc--Scutum--EGRS & $26^{\circ}<l<32^{\circ}$   & 13.5--E      &  850--E     &   92       &     --                  &  cw     &  29       & 12.5    \\
Scutum\_1               & $32^{\circ}<l<38^{\circ}$   & 4.0--8.0          &  200--580        &   73     &  1.2$\pm$0.6          &  ccw   & 35    & 6.0\\
Scutum\_2               & $32^{\circ}<l<38^{\circ}$   & 8.0--15.0          & 540--900       &   19       &  -2.0$\pm$0.9          &  cw   & 35     & 9.0 \\
Scutum--EGRS             & $32^{\circ}<l<38^{\circ}$   & 15.0--E          & 900--E       &   87       &  --                    &  --     & --     & -- \\
Scutum-Sgr               & $38^{\circ}<l<44^{\circ}$   & 6.0--12.0         & 200--500        &   52      &  3.0$\pm$0.9          &  ccw       & 41     & 9.0 \\
Scutum-Sgr--EGRS         & $38^{\circ}<l<44^{\circ}$   & 12.0--E      & 500--E        &   105      &    --                   &  cw        & 41     & 13.0 \\
inner Sagittarius              & $44^{\circ}<l<50^{\circ}$   & 3.0--12.0          & 100--350        &   38      &  3.9$\pm$1.3         &  ccw        & 47     & 7.5 \\
inner Sagittarius--EGRS        & $44^{\circ}<l<50^{\circ}$   & 12.0--E     & 350--E        &   148      &   --                    &  cw        &  47       & 12.5 \\
mid Sagittarius              & $50^{\circ}<l<56^{\circ}$   & 8.5--12.0        & 300--600       &   13       &  1.5$\pm$0.6          &  ccw        & 53     & 10.0 \\
mid Sagittarius--EGRS        & $50^{\circ}<l<56^{\circ}$   & 12.0--E        & 600--E       &   126       &  --                &  cw          &  53       & 12.0 \\
outer Sagittarius              & $56^{\circ}<l<62^{\circ}$   & 3.0--8.5          & 40--300         &   22       &  0.7$\pm$0.3          &  ccw      & 59     & 5.0 \\
outer Sagittarius--EGRS        & $56^{\circ}<l<62^{\circ}$   &  8.5--E      & 300--E         &  184       &  --                  &  cw          &   59      & 8.0 \\
Local-Perseus\_1         & $62^{\circ}<l<70^{\circ}$   & 0--6.0          &  0--120          &   14      &  -1.9$\pm$0.7          &  cw       & 66     & 4.0 \\
Local-Perseus\_2         & $62^{\circ}<l<70^{\circ}$   & 6.0--9.0          &  120--300        &   22      &  1.4$\pm$0.6          &  ccw       & 66     & 7.5  \\
Local-Perseus--EGRS       & $62^{\circ}<l<70^{\circ}$   & 9.0--E          &  300--E        &   232      &   --                 &  cw         &  66     & 10.5 \\
Local-Perseus             & $70^{\circ}<l<80^{\circ}$   &   --              &     --          &   --         &   --        &  --      &  --    & -- \\
Local-Perseus--EGRS       & $70^{\circ}<l<80^{\circ}$   &  --               &   --            &   --      &     --           &  --          &      --   &  --\\
Local-Perseus             & $80^{\circ}<l<90^{\circ}$   & 1.0--5.0          & 20--200         &   11       &  -2.8$\pm$0.6          &  cw       & 85     & 3.0 \\
Local-Perseus--EGRS       & $80^{\circ}<l<90^{\circ}$   & 5.0--E          & 200--E         &   244       &  --              &  ccw         &    85        & 5.5 \\
\hline
%\multicolumn{9}{l}{cw means clockwise and ccw means counterclockwise.} \\
%\multicolumn{9}{l}{The field direction obtained by arm to EGRS is indicative and inferred from RMs of the most distant pulsars and the }\\
%\multicolumn{9}{l}{median RM of EGRS.}
\end{tabular}
\begin{tablenotes}
\item  cw means clockwise and ccw means counterclockwise. The field direction obtained by arm to EGRS is indicative and inferred from RMs of the most distant pulsars and the median RM of EGRS.
\end{tablenotes}
\end{table*}

\citet{hmvd18} have shown counterclockwise magnetic fields in spiral arms from pulsars with positive RMs and reversed field directions in interarm regions inferred from distant pulsars and EGRS. However, magnetic fields beyond the tangential region of spiral arms are not well constrained yet. We make quantitative analysis of RM variation against DM and distance for pulsars in the first quadrant following \citet{hmvd18}. Similar to the halo pulsars, we make the robust straight-line fitting \citep[see \S15.7 in][]{ptvf92} to RMs against distance over specified distance intervals and RM against DM in DM ranges which match the distance range as closely as possible. The longitude intervals are divided equally with a width of 6$^{\circ}$ at $26^{\circ}<l<62^{\circ}$ but a bit wider at $l>62^{\circ}$ due to sparse data at larger longitudes.  The specified distance intervals for each longitude range are guided by arm/interarm zones characterized by the spiral model of \citet{hh14}. 
Again, pulsar RMs with uncertainties larger than 30 rad m$^{-2}$ are omitted. The regions for which we have analyzed the RMs are listed in Table~\ref{BdiskQ1}.

We derive the magnetic field directions from the RM trends beyond the tangential regions of spiral arms that have not been detected before (see Figure~\ref{mfdirection}). Although distances to individual pulsars may be subject to some unpredictable uncertainties, the RM distribution trends versus distance or DM for a group of pulsars can indicate magnetic fields in the range. The pulsar RMs as functions of distance and DM of the regions are presented in Figure~\ref{rmdm4q1}. The derived field directions are illustrated in Figure~\ref{mfdirection}. Locations of direction arrows at the mean longitude and the approximate mean distance for the relevant RM-DM fit are listed in Table~\ref{BdiskQ1}.

Inside the Sagittarius arm, spiral arms are tightly enwinded. Previously, the magnetic fields in the Norma arm, the Crux-Scutum arm and the Carina-Sagittarius arm  conform to the counterclockwise pattern both in the first and fourth quadrant \citep{njkk08,hml+06,hmvd18}. Evidence for clockwise fields in interarm regions is demonstrated in the fourth quadrant \citep{nk10,hml+06,hmvd18}. The clockwise field near or beyond the Scutum arm was suggested by \citet{rl94} based on negative pulsars RMs, and supported by variation of EGRS RMs behind the Galactic disk \citep[e.g.][]{vbs+11}. The careful analysis gives only weak evidence for clockwise fields in the Scutum-Sagittarius interarm, mainly based on the RM difference between distant pulsars and EGRS \citep{hmvd18}. 

Figure~\ref{rmdm4q1} shows positive RMs decreasing with distance and DM in the Near 3kpc-Scutum interarm tangential region ($26^{\circ}<l<32^{\circ}$) between 4.5 kpc and 8 kpc, conforming to the clockwise magnetic fields in the Crux-Norma interarm region previously determined in the fourth quadrant \citep{hmvd18}. Beyond the interarm region, a positive slope of more distant pulsars ($D>$ 8 kpc)  indicates a counterclockwise field that could result from the Scutum arm. 
Comparison of RMs for most distant pulsars with EGRS RMs may show another reverse farther than these pulsars, though it is not possible to determine exactly where the reversal occurs. Because both the magnetic field strength and the electron density generally decline as the Galactocentric radius increases, it is reasonable to assume the reversal happens in the next arm/interarm region.

In the Scutum tangential region ($32^{\circ}<l<38^{\circ}$) between 4 kpc and 8 kpc, the increase of RM with distance and DM provides evidence for counterclockwise fields in this region, which is already obtained by \citet{hmvd18}. The decreasing RM-DM relation beyond the tangential zone based on the newly determined pulsar RMs at $D>$ 8 kpc clearly shows a reversed field in the Scutum-Sgr interarm region or beyond. 

Pulsar RMs in the longitude intervals from $38^{\circ}<l<62^{\circ}$ further show evidence for counterclockwise pattern in different sections of Sagittarius arm and Perseus arm (farther part beyond the tangential regions of these arms), and possible field reversals between the end of fitting regions and the edge of the Galaxy by comparing RMs of distant pulsars and EGRS RMs.

The clockwise magnetic field in the Local interarm within 3 kpc between the Sagittarius arm and Perseus arm has long been known \citep[e.g.][]{man74} based on primarily negative pulsar RMs within $30^{\circ}$ of $l=90^{\circ}$ and mostly positive RMs in the opposite direction. The field direction in the Local-Perseus interarm beyond 3 kpc was not known due to insufficient pulsar data nearer than 6~kpc in the direction of $60^{\circ}<l<90^{\circ}$. The fields here also cannot be constrained by EGRS RMs behind the disk \citep{vbs+11}.
\citet{hmvd18} show clear evidence for counterclockwise fields in  the Perseus arm presented by the increasing RM pattern for pulsars at $D>$ 5~kpc in the range of $60^{\circ}<l<80^{\circ}$. They also show weak evidence for another field reversal beyond the Perseus arm based on smaller RMs of EGRS in this direction. Based on primarily new RM data from the FAST, as shown in Figure~\ref{rmdm4q1}, RMs decrease nearer than 6~kpc in the longitude interval $62^{\circ}<l<70^{\circ}$ and then RM values become positive for distant pulsars beyond 6~kpc. It implies a clockwise field in the Local-Perseus interarm region and a reversal occurs in the Perseus arm. The counterclockwise field in the Perseus arm is echoed by the neighboring longitude interval $70^{\circ}<l<80^{\circ}$ where a possible positive slope of RM-Distance relation 
is found for pulsars between 5 kpc and 9 kpc. Beyond the fitting region in the two longitude intervals, there seems to be another field reversal exterior to the Perseus arm indicated by smaller RMs of distant pulsars and EGRS RMs. Pulsar RMs in the longitude range $80^{\circ}<l<90^{\circ}$ also support the clockwise field in the Local-Perseus interarm region.

\section{Conclusions}
\label{conclusion}

Observations of Faraday rotation of a large number of pulsars inside the Milky Way give key probes to reveal the large-scale Galactic magnetic field structure. The super sensitivity of the FAST enables us to not only discover faint pulsars but also measure their polarization and hence the RMs, which were not possible before by using other radio telescopes. The obtained RMs enable us to explore the feature of the Galactic magnetic fields in much wide regions. 

We have measured rotation measures for 134 weak pulsars in the Galactic halo using the FAST, most located in the first Galactic quadrant. The basic parameters for 15 pulsars including positions and period are also improved. Together with many newly determined RMs for newly-discovered pulsars and faint pulsars, and also newly measured RMs for a large number of the known pulsars, we make tomographic analysis of the magnetic field structure both in the Galactic halo and the Galactic disk.

By analyzing the pulsar RM variations against pulsar distance and DM within a number of regions in the Galactic halo, we confirm the halo toroidal fields with opposite directions above and below the Galactic plane and determine the field strength of around 2~$\mu$G. The scale height of halo fields is estimated to be at least 2.7$\pm$0.3 kpc.

Tomographic analysis for pulsar RMs in the Galactic longitude range of 
$26^{\circ}<l<90^{\circ}$ gives evidence for the clockwise magnetic fields in two interarm regions interior to the Scutum arm and between the Scutum and the Sagittarius arm, respectively. We confirm the counterclockwise field pattern in the Sagittarius arm and Perseus arm. The clockwise field direction in the Local-Perseus interarm region and field reversals in the Perseus arm and beyond are also well revealed by newly observed pulsar RMs by the FAST.

{\footnotesize {\bf Acknowledgements.} 
This work is supported by the National Natural Science Foundation of China (Grant Nos. 11988101, 11833009, and U2031115), the Key Research Program of the Chinese Academy of Sciences (Grant No. QYZDJSSW- SLH021), and the National SKA Program of China (Grant No. 2022SKA0120103). Pengfei Wang was supported by the National Natural Science Foundation of China (Grant No. 11873058 and 12133004), and the National SKA Program of China (Grant No. 2020SKA0120200). We thank Prof. R.T. Gangadhara for careful English corrections. The GPPS survey is one of the five key projects carried out by FAST, which is a Chinese national mega-science facility built and operated by the National Astronomical Observatories, Chinese Academy of Sciences.}

%%%%%%%%%%%%%%%%%%%%%%%%%%%%%%%%%%%%%%%%%%%%%%%%%%%%%%%
%%% Acknowledgements.

%%%%%%%%%%%%%%%%%%%%%%%%%%%%%%%%%%%%%%%%%%%%%%%%%%%%%%%

%%%%%%%%%%%%%%%%%%%%%%%%%%%%%%%%%%%%%%%%%%%%%%%%%%%%%%%
%%% Conflict of interest. ????????????
%%%%%%%%%%%%%%%%%%%%%%%%%%%%%%%%%%%%%%%%%%%%%%%%%%%%%%%
%\InterestConflict{The authors declare that they have no conflict of interest.}

%%%%%%%%%%%%%%%%%%%%%%%%%%%%%%%%%%%%%%%%%%%%%%%%%%%%%%%
%%% Supplements. 
%%%%%%%%%%%%%%%%%%%%%%%%%%%%%%%%%%%%%%%%%%%%%%%%%%%%%%%
%\Supplements{}
{\footnotesize \setlength{\baselineskip}{3pt}
%%%%%%%%%%%%%%%%%%%%%%%%%%%%%%%%%%%%%%%%%%%%%%%%%%%%%%%
%%%%%%%%%%%%%%%%%%%%%%%%%%%%%%%%%%%%%%%%%%%%%%%%%%%%%%%
\bibliographystyle{raa}

\bibliography{ref}
}
\end{multicols}

\end{document}